\documentclass[letterpaper, reqno]{amsart}

\usepackage{epsfig}
\usepackage{amsfonts}
\usepackage{amssymb}
\usepackage{graphicx}

\newcommand{\Hk}{\mathcal{H}_{kin}}
\newcommand{\Hp}{\mathcal{H}_{phys}}

\newcommand{\Pro}{\hat{P}}
\newcommand{\Po}{\Pro}
\newcommand{\Ham}{\hat{H}}
\newcommand{\Hav}{\overline{H}}
\newcommand{\U}{\hat{U}}
\newcommand{\Ua}{\hat{U}_a}

\newcommand{\Proreg}{\Pro_{l,\e}}
\newcommand{\Prorega}{\Pro_{l,a}}
\newcommand{\Ureg}{\U_{l,\e}}
\newcommand{\Urega}{\U_{l,a}}

\newcommand{\ip}[2]{\langle #1, #2 \rangle}
\newcommand{\ipk}[2]{\ip{#1}{#2}_{kin}}
\newcommand{\ipp}[2]{\ip{#1}{#2}_{phys}}
\newcommand{\bra}[1]{\langle #1 |}
\newcommand{\ket}[1]{| #1 \rangle}

\newcommand{\tpsi}{\tilde{\psi}}
\newcommand{\tphi}{\tilde{\phi}}
\newcommand{\tPsi}{\tilde{\Psi}}
\newcommand{\tPhi}{\tilde{\Phi}}
\newcommand{\tq}{\tilde{q}}

\newcommand{\e}{\epsilon}
\newcommand{\ra}{\rightarrow}

\newcommand{\N}{\mathcal{N}}

\newcommand{\del}{\vartriangle}
\newcommand{\Del}{\blacktriangle}

\newcommand{\M}{\mathcal{M}}

\newcommand{\cre}{\hat{a}^\dag}
\newcommand{\ann}{\hat{a}}

\newcommand{\Ord}{\mathcal{O}}

\newcommand{\Man}{\mathcal{M}}
\newcommand{\man}{\mathcal{M}}

\newcommand{\bd}{\del}

\newcommand{\m}{\hat{m}}

\newcommand{\EV}{\hat{S}}

\newcommand{\vj}{\vec{\jmath}}

\newcommand{\be}{\mathbf{e}}
\newcommand{\bA}{\mathbf{A}}
\newcommand{\bB}{\mathbf{B}}

\begin{document}

\title[Relating Covariant and Canonical Approaches...]{Relating Covariant 
and Canonical Approaches to Triangulated
Models of Quantum Gravity}

\author{Matthias Arnsdorf}
\address{Niels Bohr Institute \\ Blegdamsvej 17, DK-2100 Copenhagen \O}
\email{arnsdorf@nbi.dk}
\date{October 2001}

\begin{abstract}

In this paper explore the relation between covariant and canonical
approaches to quantum gravity and $BF$ theory. We will focus on the
dynamical triangulation and spin-foam models, which have in common
that they can be defined in terms of sums over space-time triangulations.
Our aim is to show how we can recover these covariant models from a
canonical framework by providing two regularisations of the projector
onto the kernel of the Hamiltonian constraint.  This link is important
for the understanding of the dynamics of quantum gravity. In
particular, we will see how in the simplest dynamical triangulations
model we can recover the Hamiltonian constraint via our definition of
the projector. Our discussion of spin-foam models will show how the
elementary spin-network moves in loop quantum gravity, which were
originally assumed to describe the Hamiltonian constraint action, are
in fact related to the time-evolution generated by the constraint.
We also show that the Immirzi
parameter is important for the understanding of a continuum limit of
the theory.

\end{abstract}

\maketitle

\section{Introduction}

General relativity can be quantised by canonical or covariant methods
depending on whether one regards space or space-time as more
fundamental. Because of the incomplete nature of the subject the
relation between the two approaches is not entirely clear, even though 
one expects that in  appropriate circumstances they should be equivalent.
The broad aim of this paper is to contribute to this discussion.

General relativity is a completely constrained system and hence quantum
gravity  distinguishes itself from ordinary field theory in
that the Hamiltonian vanishes weakly. This implies for a canonical
formalism that
there is no propagator, since there is no external time variable
with respect to which it could be defined. Instead the central object
is  the  generalised projector onto the
solution space of the Hamiltonian constraint $\Ham(x)$. Formally, this
projector can be written as:
\[
\Pro = \int [dT(x)] e^{-i\int_\Sigma d^nx \Ham(x)T(x)},
\]
where $[dT(x)]$ is a  measure on the space of scalar fields
$T(x)$ and $\Sigma$ is a spatial slice of dimension $n$ of the
space-time manifold $\M = \Sigma \times \Bbb{R}$. 
This operator then defines the physical Hilbert space of the theory,
from which we have to extract all physical predictions by the
construction of suitable observables.

This projector can be expressed as
a phase space path integral in an analogous way as is done for the
propagator in ordinary field theory. 
Hence, we expect this projector to be the link between the above
canonical description and covariant approaches to the quantisation of
gravity.

Indeed, the starting point for the covariant approach is the expression
\[
Z(g_1,g_2) := \int_{g_1}^{g_2} [d{\bf g}]e^{-iS[{\bf g}]}
\]
which describes a sum over all space-time geometries ${\bf g}$ on the
cylinder $\Sigma \times [0,1]$ with boundary geometries $g_1$ and
$g_2$.
Two promising approaches towards making sense of this expression are the
dynamical triangulation and the spin-foam models.
These have in common that they they replace the above sum over
geometries by a sum over space-time triangulations.

The aim of this paper will be to show that the projector $\Pro$  can be
suitably regularised so that it describes a sum over space-time
triangulations. More precisely, we will be discussing canonical
theories of quantum gravity that have a Hilbert space of kinematical
states $\del$ given by triangulations or dual graphs of the spatial
slice $\Sigma$. 
We will provide two different regularisations of the projector such that
its matrix elements
$\bra{\del_1}\Pro\ket{\del_0}$ are given by a sum over triangulations of
the cylinder $\Sigma \times [0,1]$, with boundary triangulations
$\del_0$ and $\del_1$. The amplitudes of the triangulations will be
given by the dynamical triangulation and
spin-foam models respectively.

The idea will be to decompose the projector into a product of
``small'' unitary local proper time evolutions. Acting on an initial
triangulation $\del$ each elementary evolution corresponds to a gluing
of an $(n+1)$-dimensional simplex to the boundary. In this way we build
up a sum over space-time triangulations with the correct amplitude.

This link between canonical and covariant theories is
important for the understanding of the dynamics of these quantum
gravity models and for providing a space-time picture in the canonical 
framework. Ultimately, one would like to relate a
given Hamiltonian to the amplitude of a corresponding covariant
model. As we will see this can be achieved in the simplest case
of 1+1 dynamical triangulations.
The discussion of  spin foam models of quantum gravity and the closely
related topological $BF$ theories, will clarify their relation to 
canonical loop quantum gravity. In particular, we will see that
elementary moves of spin-networks, which were thought to describe
the action of the Hamiltonian constraint, are in fact
related to time-evolutions. A better understanding of this
relation should help in deciding which of the 
current proposals for spin
foam models provides a correct description of quantum gravity and
also shed light on the nature of the Hamiltonian constraint of loop
quantum gravity.

This work builds on a previous attempt to derive a spin foam model
from 3+1 loop quantum gravity by Reisenberger and
Rovelli~\cite{Reisenberger:1997pu}, 
which was developed further by Rovelli in~\cite{Rovelli:1998dx}.
 Our approach will
differ in 
 the regularisation of the projector. We will see in this paper how
this leads to many attractive features:  
\begin{itemize}

\item By providing two different regularisations we can describe both
the spin-foam and the dynamical triangulation models.

\item Our expansions are closely related to the derivation of the
path integral in quantum mechanics and field theory.

\item They provide a clear  role for the $[dT(x)]$ integration over scalar
(lapse) functions in the projector and avoid infinities due to this integration.

\item They provide a very natural interpretation of the addition
of space-time simplices as local proper time evolutions.

\item They avoid  over counting of space-time
triangulations.

\item We obtain an understanding of the continuum limit of spin-foam 
models.

\item Most importantly one can show that in $BF$ theories and
2+1 loop quantum gravity the addition of tetrahedra, when viewed
as an operation on physical states, leaves these invariant. This
suggests that these actions should be viewed as evolutions and
not actions of the Hamiltonian constraint operator.

\item In addition, in 1+1 dynamical triangulation we can show
explicitly how  we can derive the correct Hamiltonian operator
from the covariant model via our expansion of the projector.

\end{itemize}

We begin our discussion by describing the formal relation between the
projector and path integrals for quantum gravity.  This will be made
concrete in the dynamical triangulations framework focusing on 1+1
dimensions, where results can be stated most explicitly and simply. We
then continue to generalise our results to higher dimensions before
moving on to consider spin-foam models and loop quantum gravity.

\section{Canonical vs. Covariant}

In this section we describe the generalised projector and show how it
is formally related to a sum over geometries. We will find that
canonical and covariant quantum gravity are equivalent up to the
choice of range in the integration over lapse functions.

\subsection{Canonical quantum gravity}

The canonical description of general relativity requires that we
restrict our attention to space-time manifolds $\M$ that allow a split
into space and time, i.e.\ manifolds diffeomorphic to $\M = \Sigma
\times \Bbb{R}$.  Here and in the following $\Sigma$ will denote the
$n$-dimensional spatial slice, which we will take to be compact.

 General relativity is a completely constrained
system, i.e.\ the Hamiltonian for the theory vanishes weakly. This
can be viewed as a consequence of diffeomorphism invariance and
the absence of an external time parameter. Hence  quantisation
follows the Dirac prescription which requires that we quantise
the unconstrained classical system and impose the constraints as
quantum operators on an initial, \emph{kinematical}, Hilbert
space $\Hk$. Loosely speaking, the constraints implement the
diffeomorphism invariance of the theory. 
Typically, we distinguish between the ``momentum'' constraints $H_i$, 
which generate diffeomorphisms of the spatial slice $\Sigma$ and the
``Hamiltonian'' constraint $H$,
which is responsible for
diffeomorphisms normal to the spatial slice and is hence
associated with dynamics. If connection instead of metric variables
are used in the coordinatisation of the phase space there will be
additional ``Gauss'' constraints, enforcing invariance under internal
gauge transformations corresponding to the rotation of tetrads.

Physical states lie in the kernel of the constraint operators. If this
space is non-trivial and can be endowed with an inner product, we
obtain the final \emph{physical} Hilbert space $\Hp$.  Note the
absence of a Hamiltonian operator on $\Hp$, resulting in the
the lack  of a notion of evolution or propagator. States are interpreted
as entire histories, which makes the extraction of physical
information from this frozen picture highly problematic.

These conceptual matters will not concern us. Instead we need to
face the following technical issue related to our problem of
constructing a path integral for general relativity. Whenever
0 lies in the continuous part of the spectrum of one of the
constraint operators, the corresponding eigenvectors will not be
normalisable. Hence the construction of the physical Hilbert
space is non-trivial and proceeds by the so-called refined
algebraic quantisation (RAQ) procedure (see~\cite{Marolf:2000iq} for a review
and references).  The idea is
to look for distributional solutions to the constraints, i.e.\
functionals on the kinematical state space\footnote{To be precise
we need to consider a dense subspace of the kinematical Hilbert
space. This technical issue will not be important for our formal
discussion.}.  The solutions are precisely those functionals that
are constant on the orbits of the unitary action generated by the
constraints. Equivalently we could be view the orbits
themselves as the generalised solutions. The key step is to define an
inner product on these physical states, which is done by
introducing a generalised projector $P$. This is a map from $\Hk$
onto the space of functionals, which maps a vector to its orbit under
the constraint action. The image of the map $P$ will be referred to as 
the solution space of the constraints.

Let us denote physical states by a tilde, i.e.\ $\tpsi :=
P\psi$, $\tphi := P\phi$ with $\psi,\phi \in \Hk$,
 then the physical
inner product is given by:
\begin{equation}
\ipp{\tpsi}{\tphi} := \tphi[\psi],
\end{equation}
where the right hand side denotes the action of the functional $\tphi$
on the kinematical state $\psi$.
In many cases $P$ can be constructed by the method of group
averaging. In this case we have:
\begin{equation}
\ipp{\psi}{\phi}  := \int dg \ipk{\psi}{\U(g)\phi} := \ipk{\psi}{\Pro\phi}
\end{equation}
where $\U(g)$ denotes the group action generated by the constraint
under consideration and $dg$ is a suitable measure on
the group. We have also made use of the kinematical inner product
$\ipk{\cdot}{\cdot}$ on $\Hk$.

 Completion of the solution space with respect to the
physical  inner product gives us the physical Hilbert space $\Hp$.
The operator $\Pro$ defined above is called the generalised
projection operator since it reduces to an ordinary projection
 onto the kernel of the constraints if the corresponding
group action is compact. But the power of the approach lies in
the fact that it can be applied in other cases also, in fact
whenever the above integral converges, group averaging defines the unique
generalised projector~\cite{Giulini:1999kc}.

When applied to the momentum and Gauss constraints the above procedure
is implemented simply, by demanding that states are invariant under 
the action of spatial diffeomorphisms and gauge transformations
respectively.
The Hamiltonian constraint, however, causes difficulties since it
generates diffeomorphisms which are normal to the spatial slice and
hence does not have a simple geometrical action on kinematical states.
 Indeed, the Hamiltonian constraint is related to evolution in
coordinate time hence we expect the projector corresponding to $H$ to
be related to the propagator in field theory. This is what we explore
in the following.

In analogy to the group averaging procedure we can write down a formal
expression for the generalised projection operator corresponding to the
Hamiltonian constraint operator $\Ham(x)$:
\begin{equation} \label{projector}
\Pro = \int  [dT(x)] e^{-i\Ham[T]},
\end{equation}
where $\Ham[T] := \int_\Sigma d^nx \Ham(x)T(x)$. Here $[dT(x)]$ denotes a
 measure on the space of scalar functions $T(x)$, which we leave 
unspecified. The integral should be understood as an integral over the 
generators of the algebra generated by $\Ham(x)$. Since the
exponential $\exp(-i\Ham[T])$ is linear in $T(x)$, the above can also 
be viewed as a expression of a product of delta functions
imposing the constraint $\prod_{x \in \Sigma} \delta[\Ham(x)]$.
Alternatively, we can understand the above in the following way.  The
expression $\ipk{\psi}{e^{i\Ham[T]}\phi}$ defines the ``propagator'' in
multi-fingered proper time $T(x)$ between two kinematical states
$\psi$ and $\phi$, which describes geometries on the boundaries
of the cylinder $[0,1] \times \Sigma$. Since physical results
should be independent of the arbitrary proper time separation between
initial and final splices, we should integrate over all possible
$T(x)$. This then defines the above projector. A more in depth
discussion is given in~\cite{Reisenberger:1997pu}.
We will now show in more detail how the projector and the path
integral for quantum gravity are related.

\subsection{Covariant quantum gravity}

The projector can be expressed as a phase space path integral as
is demonstrated by Klauder~\cite{Klauder:2000gu} for general constrained
systems. This phase space integral can also be derived from a sum over 
geometries approach to quantum gravity, as has been investigated in
detail by Teitelboim~\cite{Teitelboim:1982ua,Teitelboim:1983fk},
 forging the link between
covariant and canonical frameworks.

Let us denote the points in the phase space of general relativity as
$[p(x),q(x)]$, $x \in \Sigma$ and let us furthermore assume we have
eigenstates $\ket{q} \in
\Hk$ of the corresponding
configuration operators. We can then
express the physical inner product:
\begin{equation}
\ipp{\tq_1}{\tq_0} := \int
[dT(x)] \bra{q_1} e^{-i\Ham[T]} \ket{q_0}
\end{equation}
as a path integral in complete analogy with the derivation of the path
integral in elementary quantum mechanics. More concretely, we imagine
that the operator $\exp(-i\Ham[T])$ is evolving between a state describing
an initial geometry at coordinate time 0 and a final geometry at
coordinate time 1. We then write:
\begin{equation}
e^{-i\Ham[T]} = \lim_{\e \ra 0} \prod_{k=1}^N e^{-i\e \Ham[T]},
\end{equation}
with $\e = 1/N$. Then using the standard techniques we obtain formally:
\begin{equation} \label{pro-soh}
\ipp{\tq_1}{\tq_0} =\int [dT(x)] \int_{q_0}^{q_1} [Dp][Dq]
e^{i\int_0^1 dt\int d^nx (p(x)_t\dot{q}(x)_t -   H(p,q) T(x))}
\end{equation}
The expression on the right is an integral over all paths in phase
space
\begin{equation}
 t \ra [q(x)_t,p(x)_t] \hspace{1cm} t \in [0,1]
\end{equation}
 with fixed initial and final
configurations $q_0$ and $q_1$. The measure $[Dq] := \prod_{t=0}^1 [dq(x)]$
is a product of suitably normalised  measures $[dq(x)]$ on the space
of functions $q(x)$ with a  similar expression for $[Dp]$.

This is the expression that is obtained if we start from a sum
over histories approach to quantum gravity modulo ambiguities
concerning the integration range in the $[dT]$ integral. 
There one is interested in
defining the quantity:
\begin{equation}\label{soh}
\int_{g_0}^{g_1} [d{\bf g}] e^{-iS[{\bf g}]}
\end{equation}
where $S$ denotes the action for general relativity. The integral
is over all space-time geometries\footnote{Here and in the following bold
face characters will correspond to space-time quantities}  ${\bf
g}$ that interpolate between initial and final geometries $g_0,g_1$
on the boundary of the cylinder $[0,1] \times \Sigma$. By a
geometry we mean an  equivalence classes of metrics under
diffeomorphisms.

This sum over geometries can be reexpressed as a phase space path
integral. In metric variables the phase space coordinates are given by
the metrics $g_{ik}(x)$ on the slice $\Sigma$ and their conjugate
momenta $p^{ik}(x)$.

 By choosing a foliation of the space-time cylinder we can rewrite the
gravitational action in Hamiltonian form. The action is then a
function of the paths $t
\ra [g_{ik}(x,t), p^{ik}(x,t)]$, where the pair $[g_{ik}(x,t),
p^{ik}(x,t)]$ describes the geometry of the spatial slice $\{t\}
\times \Sigma$
\begin{equation}
S[N,N^i,p,g] = \int_{0}^{1} dt \int_\Sigma d^n
x(p^{ik}\dot{g}_{ik}- N(x,t)H(g_{ik},p^{ik}) - N^i(x,t)H_i(g_{ik},p^{ik}))
\end{equation}
where $N$ and $N_i$ are the lapse and shift functions which encode the
information on how the foliation of space-time is chosen.

The constraints generate gauge transformations that relate physically
indistinguishable variables. Hence, to avoid multiple counting of
equivalent paths in the  integral~(\ref{soh}) we need to fix a gauge. One of
the simplest possible gauge choices, as described by
Teitelboim~\cite{Teitelboim:1983fk}, 
is the proper time gauge $\dot{N} = 0$, $N^i = 0$. 
 These conditions fix the available
freedom up to diffeomorphisms of the initial or final slices. Hence
one can rewrite the integral~(\ref{soh}) as an integral over paths in
phase space with initial and final \emph{geometries} $g(x,0)$ and $g(x,1)$
kept fixed:
   \begin{equation}\label{soh2}
\int [dT(x)] \prod_{t=0}^1[dp^{ik}(x,t)][dg_{ik}(x,t)]e^{i\int_0^1 dt \int_\Sigma d^n
x(p^{ik}\dot{g}_{ik}- T(x)H(g_{ik},p^{ik}))}
\end{equation}
where the integral is over the functions $T(x) :=N(x,t)$, which give
the multi-fingered proper time separation between initial and final
spatial slices. 
This integral has the same form as the one derived from the canonical
projector in equation~(\ref{pro-soh}).
Note however, that
the integration measures need to be chosen in such a
way that the above integral becomes independent on our choice of
gauge, otherwise one needs to include ghosts terms in the above action
as described by Teitelboim~\cite{Teitelboim:1982ua}.
 Alternatively, by choosing the
appropriate measures Klauder has derived the above directly, without
the prior imposition of the gauge conditions~\cite{Klauder:2000gu}.  We will
not be concerned with these issues as the measures will be  implicitly determined by
the choice of models considered later, which provide concrete
realisations of the formal path-integral expressions given in this
section.

\subsection{Relating canonical and covariant approaches}

We are interested in the range of integration of the $[dT(x)]$
integral which we have left unspecified and which is responsible for a
subtlety in the relation between canonical and covariant
approaches. First we note that in order for the lapse function to arise from a well
defined foliation it is necessary that $T(x) \neq 0$ for all
$x$. Since $T(x)$ is a continuous function this implies that $T(x)$ is
either entirely positive or entirely negative. 

If the
integral~(\ref{soh2}) is to project onto the kernel of the Hamiltonian
constraint and be equivalent to the expression~(\ref{pro-soh}) then 
we need integrate over both positive and negative lapse
functions, as is clear from the analogy between the projector and a
product of delta functions.

However, if we want the phase space integral~(\ref{soh2}) to
be equivalent to the sum over geometries integral~(\ref{soh}), then we
should only integrate over half of the allowed $T(x)$, say only those
for which $T(x)>0$. 
To see this we recall that that the function $T(x)$
encodes how the space-time manifold $\M$ is foliated as $[0,1] \times
\Sigma$. In equation~(\ref{soh2}) we are integrating over the
variables $[T(x),g_{ik},p^{ik}]$. The function $T(x)$ determines how
the 3-metric on each slice $\{t\}\times\Sigma$ and their conjugate momenta
are glued together to give a space-time metric on $\M$. However, a change in
sign of the proper time $T(x)$ simply corresponds to reversing the foliation
of $\M$ and hence  $[T(x),g_{ik},p^{ik}]$ and $[-T(x),g_{ik},p^{ik}]$
will describe equivalent space-time metrics.

This symmetry is not enforced by the constraints as they only identify
configurations that can be related by infinitesimal
deformations. Hence the sum over geometries expression is not exactly equivalent
to the phase space integral derived from the canonical picture 
due to a larger symmetry group
in the sum over geometries case. This  is reflected in the range of
integration in the integral\footnote{Teitelboim  gives an argument 
why one might want to restrict the range of integration in the
canonical case also~\cite{Teitelboim:1983fh}. 
The idea is that one should implement
causality by demanding that one integrates only over configurations
where the final slice is to the future of the initial one.
}~(\ref{soh2}). 

This feature can be seen clearly in the discussion of the relativistic
particle, which can be viewed as general relativity in 0+1
dimensions. If we choose the half infinite range of integration then
the phase space path integral will define the Feynman propagator (an in
depth discussion of the different Greens functions that can be
obtained in this approach is given by Halliwell and Ortiz
in~\cite{Halliwell:1992nj}). 
This is the same result that is obtained if we try to define the sum
over geometries integral, which is just a function of the length of
the particle trajectory. One approach to do this is to use a dynamical
triangulations approach in the Euclidean regime and then Wick rotate
the result, which gives us precisely the Feynman propagator.

Let us summarise. We have seen that $\Pro$ defined in
equation~(\ref{projector})
is an  important operator for  canonical quantum gravity. It has two
interpretations depending on the $T(x)$ integration range:
\begin{enumerate}
\item For the full integration range we obtain the projector onto the
kernel of the Hamiltonian constraint;
\item For the half-infinite integration range $\Pro$ is related to the sum 
over geometries.
\end{enumerate}
In either case $\Pro$ can be described by a path integral. 

In the
following we will show how this can be made concrete by discussing
first the dynamical triangulations and second the spin-foam model of
quantum gravity.
The former gives a definition of a sum over geometries whereas the
latter is a path-integral definition of the projector. We will see how 
these descriptions can be recovered from two respective
regularisations of $\Pro$ defined on the appropriate kinematical
Hilbert spaces.

\section{Dynamical triangulations}

The dynamical triangulations approach to quantum gravity attempts to
regularise and define the path integral given by equation~(\ref{soh})
(c.f.~\cite{Ambjorn:1999nc} for details and references). The idea is to work
with manifestly diffeomorphism invariant quantities and thus avoid
difficulties concerning gauge fixing. This is achieved by replacing
the smooth space-time metric manifold by a simplicial manifold
constructed from
\emph{equilateral} (n+1)-dimensional simplices of geodesic edge
length $a$. Here and in the following we will be working with
Riemannian metrics. A Lorentzian framework for dynamical
triangulations exists~\cite{Ambjorn:2001cv} but for our purposes the
Riemannian case will suffice to exemplify all the features of our
construction.

We are approximating the sum over geometries by a sum over
triangulations of the space-time manifold. The weight of each
triangulation is given by the Euclidean Regge action, which can be
expressed solely as a function of the squared edge lengths. But note
that in contrast to Regge calculus we are keeping the edge lengths
fixed --- triangulations differ only by the connectivity of the
simplices.  The challenge is to find a non-trivial continuum limit of
this statistical model by taking $a \ra 0$, and appropriately
rescaling the bare Newton and cosmological constants. This works well
in 1+1 dimensions, where one can recover results from standard Liouville
theory. In higher dimensions evidence for a interesting continuum
limit is still lacking.

Our goal is to show how the sum over triangulations can be recovered
from a canonical theory via an  expansion of the projector defined in
equation~(\ref{projector}). We assume
again that space-time is of the form $\mathcal{M} = \Sigma \times
\Bbb{R}$. The kinematical Hilbert $\Hk$ space for our theory is
constructed from the free vector space generated by all equilateral
triangulations $\del$ of $\Sigma$. By identifying the edges as
geodesics of length $a$ we can interpret states as piecewise-linear
geometries.

We will distinguish between auxiliary and kinematical
states. Auxiliary states are actual embedded triangulations of
$\Sigma$, whereas kinematical states are equivalence classes of
auxiliary states under diffeomorphisms, i.e.\ we can think of them as
non-embedded or abstract triangulations which are simplicial manifolds 
with the topology of $\Sigma$. 

An inner product is specified on the space of 
triangulations by demanding that inequivalent triangulations are
orthogonal i.e.:
\begin{equation}
\ip{\del}{\del'} := \delta_{\del,\del'}
\end{equation}
Completion of the space of kinematical triangulations with respect to
this inner product gives us $\Hk$.


We are interested in the projector $\Pro$ acting on this space. 
From the discussion in the last section we know that $\Pro$ is related
to the sum over geometries, which in the dynamical triangulation
approach is given by
\begin{equation}
Z_a(\del_1,\del_0) := \sum_{\Del} e^{-S[\Del]}
\end{equation}
 which denotes the sum over all space-time triangulations $\Del$
with boundaries $\del_0$ and $\del_1$ weighted
by the exponential of the Regge action $S[\Del]$.
In particular,  we expect that by restricting the integration range in the
projector we should have:
\begin{equation}\label{half-inf}
\int [dT(x)]_+ \bra{\del_1}e^{-\Ham_a[T]} \ket{\del_0} = Z_a(\del_1,\del_0)
\end{equation}
where $[dT(x)]_+$ denotes an integration over positive proper-time
functions only\footnote{Note that this restriction implies that the expression on
the left of equation~(\ref{half-inf}) does not define the physical
inner product.}.
Note that because we are doing a Riemannian calculation we are using
real amplitudes.

We now show how this can be demonstrated by choosing a suitable
regularisation of the projector

\subsection{Regularisation of the projector I}\label{expansion1}

Let us consider the projector with half-infinite integration
range\footnote{Although this is strictly speaking not a projector we
will, by abuse of notation, continue to refer to it as such.}
\begin{eqnarray}
\Pro &=& \int [dT(x)]_+ e^{-\int d^n x T(x) \Ham(x)}\\
&=& \int_0^\infty dv \int[dT(x)]_+ \delta\left[\int d^nx T(x) = v
\right] e^{-\int d^nx T(x) \Ham(x)}. 
\end{eqnarray}
 Note that the dimension of $v$
is naturally that of space-time volume. Hence we are splitting the sum
over all geometries into a sum over geometries with fixed volume and
then summing over all possible volumes.

To regularise the functional integral, we allow the function $T(x)$ to
only take values in discrete steps of $\e$. In other words $T(x) =
t(x)\e$ with $t(x) \in \Bbb{N}$. We also write the spatial integrals
as Riemann sums. To do this we assume that we have given some metric
on $\Sigma$ with respect to which we can define a square lattice with
lattice spacing $l$. We will denote the set of
vertices of this lattice by $V$. This gives us a regulated projector
$\Proreg$.
\begin{eqnarray}
\Proreg &=& \int_0^{\infty} dv \int [dT]_\e \delta \left[ l^n\e \sum_{x
\in V} t(x) = v \right] e^{-\sum_{x \in V} \e l^n t(x)\Ham(x)}\\
&=& \sum_{s=0}^\infty \int [dT]_\e \delta\left[\sum_{x \in V} t(x) = s
\right] \prod_{x \in V} \prod_{k=0}^{t(x)} e^{-\e l^n \Ham(x)}
\end{eqnarray}
where $[dT]_\e$ denotes a measure on the positive discretised lapse functions.

Let us set 
\begin{equation}
\Ureg(x):= e^{-\e l^n \Ham(x)}.
\end{equation}
 One can see that the above
delta function imposes the condition that the product
$\prod_{x \in V} \prod_{k=0}^{t(x)} \Ureg(x)$ has exactly $s$
factors. The integral over the discretised lapse function then ensures
that for fixed $s$  we sum over all possible $t(x)$ satisfying the
constraint. More precisely we have:
\begin{equation}\label{expansion}
\Proreg = \sum_{s=0}^\infty \sum_{\vec{x} \in V^s} \Ureg(x_1)\cdots\Ureg(x_s)
\end{equation}
where the second sum is over all possible \emph{ordered} $s$-tuples
$\vec{x}:=(x_1,...,x_s) \in V^s$. The ordering arises in the replacement
of the spatial integrals by Riemann sums. For example, in (1+1)d the
ordering implies that $x_1 \leq x_2 \leq ... \leq x_s$.

The continuum projector is recovered in the limits $\e, l \ra 0$,
whereby we need to check that the results are independent of the
lattice chosen in the regularisation. 

In the following we will see that this expansion allows a very natural
interpretation.  The operator $\Ureg(x)$ will correspond to a
local evolution of a state in proper time $\e$, which will be
identified with the addition of a space-time simplex to the
boundary triangulation. In this way the above sum will correspond to a
sum over triangulations of space-time with the number of tetrahedra in
a particular triangulation given by $s$.
In the next sections we will discuss this in detail in the context of
1+1 dimensions.

\subsection{1+1 dimensions}

In 1+1 dimensions the action for general relativity is proportional to
the volume of space-time: $S[\mathbf{g}] = \lambda \int_\M \sqrt{-\det
\mathbf{g}}$, where $\lambda$ denotes the cosmological constant.   
In the following we will be approximating the sum over geometries by a 
sum over triangulations of the cylinder $S^1 \times [0,1]$.

Space-time simplices are simply triangles with area proportional to
$a^2$. Hence the transition amplitude between two boundary states
$\del_0$ and $\del_1$ is:
\[
Z_a(\del_1,\del_0) = \sum_\Del e^{-\lambda a^2 N_\Del}
\]
where the sum is over space-time triangulations $\Del$ with the
appropriate boundaries, and $N_\Del$
is the number of triangles in $\Del$.

 Auxiliary states correspond to triangulations of $S^1$
and are thus an embedded sequence of links. They can be characterised by
the coordinates $v_i$ of the vertices between the links, i.e. they are
of the form $\ket{\vec{v}}:=\ket{v_1...v_n}$, where $n$ is the number of links.
Since the only diffeomorphism invariant
information contained in the auxiliary states is the number of links,
kinematical states can be though of an abstract chain of links denoted
by $\ket{n}$. These are the sates spanning $\Hk$.

The aim is to find a to find a definition of $\Ureg(x)$ acting on
$\Hk$ such that as
$l \ra 0$ we can express $\bra{n}\Proreg\ket{m}$ as a sum over
triangulations with the correct weights.
In other words we would like to deduce the appropriate quantisation of 
the evolution $U_{l,\e}(x)$ from the path integral form of the theory.
To do this we will identify the action of the operator $\Ureg(x)$ with
the stacking of triangles.

\begin{figure}
\begin{center}
\includegraphics*[width=5in,keepaspectratio]{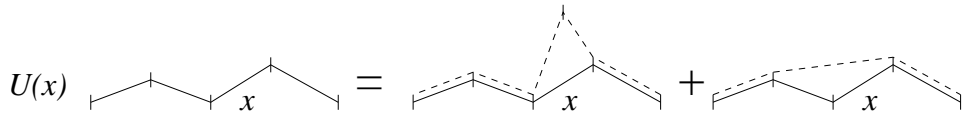}
\caption{The evolution operator acts on a kinematical state by a
adding a triangle in two possible ways. This creates a new
kinematical state which is depicted by the dashed line.}\label{action}
\end{center}
\end{figure}

There are two ways to add triangles to a sequence of links $\ket{n}$. 
 We  can glue one or two faces of the triangle to the
boundary. This creates a new set of links  $\ket{n+1}$ or $\ket{n-1}$
respectively. Roughly, we will define $\Ureg(x)$ to be the sum of both
these local actions as shown in figure~\ref{action}.
Repeated action of $\Ureg$ will then create a triangulation. An
important aspect  of this identification (which is unique to the
dynamical triangulations framework) is  that
the parameter $a$ in our  model can be
interpreted as a proper time distance. Hence it is natural to identify
$a$ with $\e$, as we will do in the following.

\begin{figure}
\begin{center}
\includegraphics*[height=1.1in,keepaspectratio]{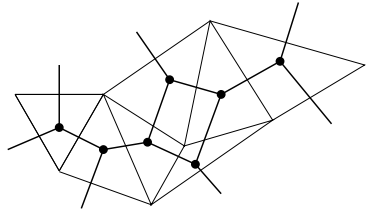}
\caption{A 2-dimensional triangulation and its dual trivalent graph}\label{dual}
\end{center}
\end{figure}

These ideas can be expressed more precisely in the dual picture, which is
also more likely to be useful in generalisations to higher dimensions
and spin-foams. In general, the dual of a $(n+1)$-dimensional triangulation is an
$(n+2)$-valent graph. The graph has a node in the centre of each simplex
and one edge crossing each face as shown in figure~\ref{dual}. In one
dimension the dual of a triangulation is just a sequence of nodes. 
Hence states that are dual to an auxiliary state $\ket{v_1,...,v_n}$
will be denoted by $\ket{\vec{x}} :=\ket{x_1,..,x_n}$
where $x_i := (v_i+v_{i+1})/2$ now is the co-ordinate of the $i$'th node which corresponds
to the centre of the $i$'th link. 

The operator $\Urega(x)$, being local,
acts on auxiliary states.  Let us introduce creation and
annihilation operators $\cre(x)$ and $\ann(x)$ acting on nodes of
auxiliary dual states:
\begin{equation}
\cre(x)\ket{x_1,...x_n} :=  \left\{ \begin{array}{ll} 
				0 & \mbox{if $x \neq x_i$} \\
			\ket{x_1,...,x_{i-1},\frac{x_{i-1} +
x_i}{2},x_i, ...,x_n} & \mbox{if $x = x_i$}
			\end{array} \right.
\end{equation}
\begin{equation}
\ann(x)\ket{x_1,...,x_n} := \left\{ \begin{array}{ll} 
				0 & \mbox{if $x \neq x_i$} \\
			\ket{x_1,...,\not x_i,...x_n} & \mbox{if $x = x_i$}
			\end{array} \right.
\end{equation}

\begin{figure}
\begin{center}
\begin{tabular}{lcr}
\includegraphics*[width=1.7in,keepaspectratio]{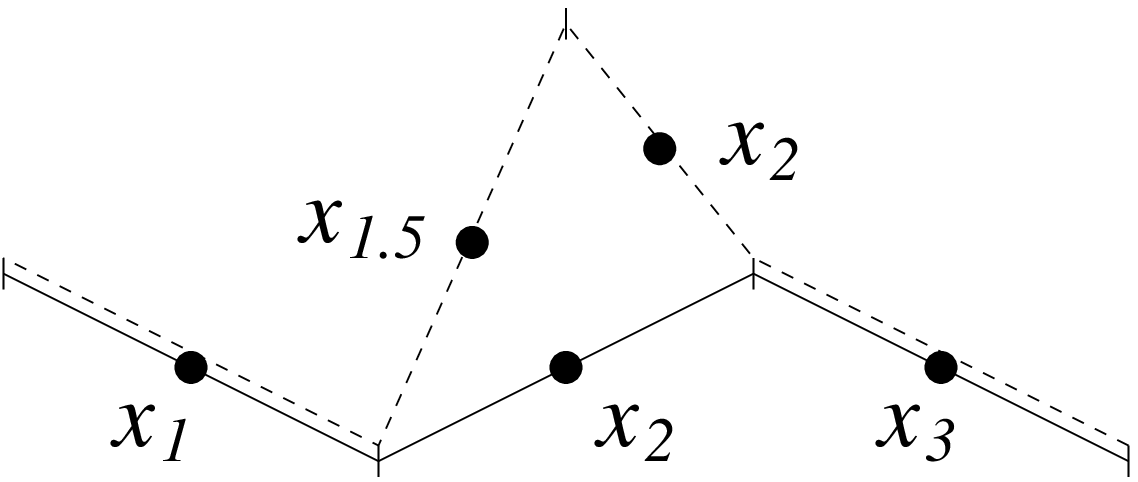} &\hspace{1.5cm}&
\includegraphics*[width=1.7in,keepaspectratio]{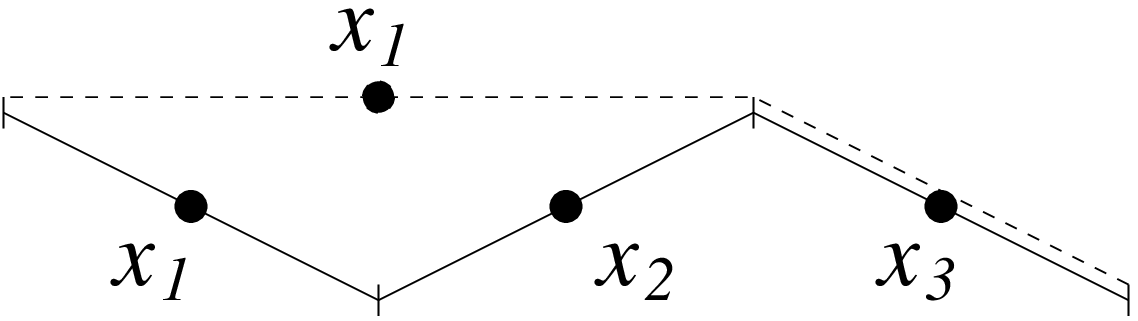}
\end{tabular}
\caption{The actions of $\cre(x_2)$ (left fig.) and $\ann(x_2)$ (right 
fig.) on
the auxiliary state $\ket{x_1,x_2,x_3}$. The newly created states are
depicted by the dashed lines and $x_{1.5} = (x_1+x_2)/2$.}\label{a}
\end{center}
\end{figure}

This can be identified with the gluing of triangles as
follows. The creation operator $\cre(x_i)$ corresponds to the gluing of
one edge of a triangle to a boundary link centred at $x_i$. 
 This gives a
new boundary triangulation, which is identified with an auxiliary
state by assigning the the node corresponding 
to right free edge of the added
triangle with the coordinate $x_i$ and the left edge with
$(x_{i-1}+x_i)/2$ as depicted on the left of figure~\ref{a}.

The annihilation operator $\ann(x_i)$ acts by gluing two edges of a
triangle to the boundary links $x_i$ and $x_{i-1}$.
This gives a new state with one less link than the original one. 
 The free edge of the added triangle  is  identified with
the node $x_{i-1}$ as can be seen on the right of figure~\ref{a}.
We use these operators to define $\Urega(x)$:
\begin{equation}\label{U-def}
\Urega(x)\ket{\vec{x}} := \frac{1}{2}e^{-a^2\lambda}(\cre(x) +
\ann(x))\ket{\vec{x}}
\end{equation}

Multiple actions of $\Urega(x)$ will thus give rise to a sum of
auxiliary states, each of which can be identified with an abstract
triangulation. 
The projector is given by a sum of ordered products
$\Urega(x_1)\cdots\Urega(x_s)$ as shown in
equation~(\ref{expansion}).
In the (1+1)d case the ordering means that $x_1 \leq x_2 \leq...\leq
x_s$. This implies that an ordered product of evolution operators will
construct triangles from right to left (if the coordinate axis points 
left to right). 

The action of $\Urega(x)$ on a state $\ket{\vec{x}}$ will
be $0$ whenever $x$ is not the coordinate of a node in $\ket{\vec{x}}$.
It follows that whenever the action of a product
$\Urega(x_1)\cdots\Urega(x_s)$ is non-zero it will act by constructing a
sum of triangulations, each containing $s$ triangles.
 Hence $\Prorega$ corresponds to a sum over
triangulations, where the parameter $s$ in equation~(\ref{expansion})
determines the number of triangles in a triangulation.
 
 As $l \ra 0$ the separation between the
allowed values for $x \in S^1$ tends to 0. 
Hence, in this limit,  any triangulation that can be constructed by an ordered
sequence  of evolution operators  will be present in the
sum over triangulations described by $\Prorega$

We now complete the definition of $\bra{n}\Prorega\ket{m}$ as
follows. We choose an auxiliary state $\ket{\vec{m}}$ which is a
representative in the class $\ket{m}$, i.e.\ any auxiliary state that
has $m$ links. The action of $\Prorega$ on this state will produce a
sum of auxiliary states:
\begin{equation}
\Prorega\ket{\vec{m}} =
 \sum_i e^{-a^2\lambda N_{\Del_i}}\ket{\vec{x}_i},
\end{equation}
where each of these states $\ket{\vec{x}_i}$ is the (future) boundary of an abstract
triangulation $\Del_i$ containing $N_{\Del_i}$ triangles as described
above.  This gives:
\begin{equation}
\bra{n}\Prorega\ket{m} = \sum_i e^{-a^2\lambda N_{\Del_i}}\ip{n}{\vec{x}_i}
\end{equation}
The inner product $\ip{n}{\vec{x_i}}$ is given by the dual action
of $\ket{n}$ on the auxiliary state $\ket{\vec{x_i}}$. This is $1$ if
if the number of links in $\ket{\vec{x}_i}$ is $n$ and 0 otherwise. 

In this way
$\bra{n} \Prorega \ket{m}$ is given by a sum over triangulations that
have two boundaries, with $n$ and $m$ links respectively. The
amplitude for each triangulation is just $e^{-a^2\lambda N_{\Del}}$,
where $N_{\Del}$ is the number of triangles in
the triangulation as required.

Given two different auxiliary states representing the same kinematical
state (i.e.\ having the same number of nodes), we can find two ordered
sequences of operators  that will produce the same abstract
sum of triangulations when acting on the respective states. 
 This is because the coordinates of the nodes
just serve to label the links and their exact value is irrelevant for
the geometrical interpretation of the action of $\Urega(x)$.  But as
noted above, in the limit that $l \ra 0$, any 
triangulation that can be constructed by an ordered sequence of
evolutions acting on a given auxiliary state, will be present in the
expansion of $\Prorega$.  Hence, in this limit, the amplitude becomes
independent of the choice of representative for the state
$\ket{m}$. For the same reason the amplitude also becomes independent
of the metric chosen in the regularisation of $\Prorega$

\begin{figure}
\begin{center}
\includegraphics*[height=1in,keepaspectratio]{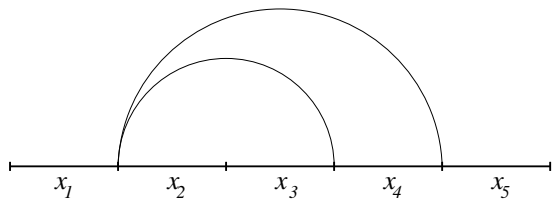}
\caption{This triangulation cannot be constructed with an ordered
sequence of $\Ureg(x)$. This is because the top
triangle has to be constructed after the bottom one. This cannot
happen since it is situated to the right. In other words we would need
a sequence $\Urega(x_4)\Urega(x_3)$, which is not allowed since $x_4
\geq x_3$.}\label{notposs}
\end{center}
\end{figure}

To conclude we note that the triangulations constructed with the above
definitions are unique, i.e.\ two different sequences of operators
$\Urega(x_1)\cdots\Urega(x_s)$ acting on the same auxiliary state will
give rise to different triangulations.  This follows essentially from
the ordering of the operators as shown in the appendix. However it is
also clear that not all triangulations are produced c.f.\
figure~\ref{notposs}. This is a feature of the choice of definition of
$\Urega(x)$. Indeed, it is possible to choose a different action of
$\Urega(x)$ in which $\ann(x_i)$ corresponds to a gluing of a triangle
to the links $x_i$ and $x_{i+1}$. In this case all triangulations are
produced albeit not uniquely.  It remains to be seen if an action can
be chosen such that we have both properties.  In this respect it will
be important to check to which universality class the continuum limit
of the triangulations model constructed above corresponds to. As we
discuss below the derivation of the Hamiltonian corresponding to this
model suggests that we are summing over geometries that suppress the
creation of baby universes.

\subsection{The Hamiltonian}

One of the benefits of working in 1+1 dimensions is that we can use
the definition of $\Urega(x)$ to derive the
form of the continuum, averaged, Hamiltonian constraint, thus
completing the link between the covariant and canonical
pictures\footnote{We thank Jan Ambj{\o}rn for this observation.}.

The Hamiltonian and $\Urega(x)$ are related via: $\Urega(x) =
e^{-al\Ham(x)}$. In the following we will take the $a$ and $l$
limits simultaneously and hence set $a = l$. Thus for small $a$ we have:
 $\Ua(x) \approx 1 -
a^2\Ham(x)$. Let us define the averaged Hamiltonian as:
\begin{equation}
\Hav_a = \int_{S^1} dx \Ham_a(x)
\end{equation}
Then 
\begin{equation}\label{U-H}
a\EV_a := \int_{S^1} dx \Ua(x) \approx \hat{L}(S^1) -a^2\Hav_a,
\end{equation}
where $L(S^1)$ denotes the length of the spatial slice $S^1$.  The
point is that the integrated evolution operator $\EV_a$ has a simple
expression.  It is the sum of the elementary evolution actions (adding
of a triangle in two possible ways) on every node of a state
$\ket{n}$.  More specifically, the matrix elements of $\EV_a$ are
given by
\begin{equation}\label{S}
\bra{n}\EV_a\ket{m} = \frac{m}{2}e^{-a^2\lambda}(\delta_{m,n-1} +
\delta_{m,n+1}) 
\end{equation}
The length $\hat{L}(S^1)$ also has a natural operator action. Since
the states $\ket{n}$ describe geometries of length $na$ it follows
that $\ket{n}$ are volume eigenstates, i.e.:
\begin{equation}\label{vol} 
V(\Sigma)\ket{n} = na\ket{n}.
\end{equation}
 Hence 
from equations (\ref{U-H}), (\ref{S}) and (\ref{vol}) we learn that the matrix
elements of the discretised averaged Hamiltonian operator should be
given by:
\begin{equation}\label{discreteH}
\bra{n}\Hav_a\ket{m} \approx  \frac{m}{a}(\delta_{m,n} -
\frac{1}{2}e^{-a^2\lambda}(\delta_{m,n-1} + \delta_{m,n+1}))
\end{equation}

The states  $\ket{n}$ describe geometries of length $N =
an$. We want to take the limit $a \ra 0$ while keeping $N$ fixed. In
this way we obtain a continuum Hilbert space spanned by the states
$\ket{N}$. We are looking for the continuum expression $\Hav$ of
$\Hav_a$ which acts on this space i.e. we are interested in:
\begin{equation}
\Hav f(N) = \int dM \bra{N}\Hav\ket{M} f(M)
\end{equation}
Since the discrete states  $\ket{n} = \ket{N/a}$,
$\ket{m} = \ket{M/a}$ become increasingly better approximations to the
continuum states $\ket{N}$ and $\ket{M}$ as $a \ra 0$, we can replace
the above by:
\begin{equation}
\Hav f(N) = \lim_{a \ra 0} \sum_m \bra{n}\Hav_a\ket{m} f(ma)
\end{equation}
Using the fact that $n =
N/a$ and $m = M/a$ together with equation~(\ref{discreteH}) we obtain 
\begin{equation}
\Hav f(N) = \lim_{a \ra 0} \frac{N}{a^2}(f(N) -
\frac{1}{2}e^{-a^2\lambda}(f(N-a) + f(N+a)))
\end{equation}
Expanding $f(N-a)$, $f(N+a)$ to second order in $a$ and also expanding
the exponential this gives
\begin{eqnarray}
\Hav f(N) &=& \lim_{a \ra 0}  \frac{N}{a^2}(f(N) -
\frac{1}{2}(1-a^2\lambda)(2f(N) + a^2\frac{d^2}{dN^2}f(N) +
\Ord(a^3))) \\
&=& N\lambda f(N) - N\frac{1}{2}\frac{d^2}{dN^2} f(N) 
\end{eqnarray}

Hence we deduce
\begin{equation}
\Hav = \lambda N  - N \frac{1}{2}\frac{d^2}{dN^2}.
\end{equation}
This is the same expression (up to a factor of 2) that has been
obtained in the study of dynamical triangulation models~\cite{Ambjorn:1999nc}.
 It corresponds to a choice of triangulation models in which baby
universe creation is not allowed\footnote{
Durhuus and Lee have shown~\cite{Durhuus:2001sp} that $\lim_{a \ra 0} \bra{N/a}e^{-T 
\Hav_a} \ket{M/a} = G(N,M;T)$. Here $G(M,N;T)$ denotes a sum over
sliced triangulations with $T$ slices as depicted in
figure~\ref{slices} (see~\cite{Ambjorn:1999nc} for details). From our perspective it 
is clear why this should be so. We have $e^{-T\Hav} =
\exp[-T\int_{S^1}dx \Ham_a(x)] = [\prod_{x \in S^1} \U_a(x)]^T$. Hence
the action of this operator corresponds to the stacking of $T$ layers
of triangles to a boundary triangulation.
}.

\subsection{Discussion}

We have succeeded in deriving the form of the averaged
Hamiltonian constraint operator from the regularised path integral for
1+1 quantum gravity using our projector expansion.
This hinged crucially on the identification of the elementary
evolutions with the addition of triangles. Moreover,
 this allowed us to naturally identify  the small proper time step $a$ 
occurring in the regularisation of the projector
 with the edge length of the triangles. This gives rise to an 
interpretation to the integral over proper times $[dT(x)]$   as a sum
over triangulations.

These features are not present in the original regularisation of the
projector proposed by Reisenberger and
Rovelli~\cite{Reisenberger:1997pu}. They suggested an expansion of the
exponential in the projector
\begin{equation}\label{rr-expansion}
\Pro = \int [dT] \sum_{t = 0}^{\infty} \frac{(-\Ham[T]^t)}{t!},
\end{equation}
 which as we will discuss later this was motivated by
the form of the Hamiltonian constraint in loop quantum gravity.
In order for this expression to be interpretable as a sum over
triangulations it turns out that we need to identify $\Ham[T]$ with
the integrated move operator $\EV_a$.  In this way multiple action of
the Hamiltonian will produce all possible triangulations.

This leaves the interpretation of the $[dT(x)]$ integral unclear
since a sum over all triangulations is already generated for every
function $T(x)$.
 Intuitively, the proper
time should be reflected in the triangulation. Indeed in the dynamical
triangulations approach it is clear that the proper time corresponds
to the number of edges of triangles along a particular path.

Additionally, a problem arises because there are many ways in which a
given triangulation can be constructed by a sequence of stacking of
triangles. All these possibilities will be included in the sum
generated by the above expansion\footnote{This is avoided in our approach
because the triangles are stacked in ordered sequences.}.
This number is a non-trivial function of the triangulation and does
not depend solely on $t$, the number of triangles in the
triangulation. Hence the sum over triangulations will contain
non-trivial symmetry factors that do not occur in the definition of
$G_a(\del,\del')$. 

Crucially, however, because one has made the identification $H[T] =
\EV_a$ it will no longer be possible to derive the correct expression
for the continuum Hamiltonian.

\subsection{Generalisations}

It is straightforward to generalise our discussion to higher
dimensions. Space-time triangles will be replaced by
$(n+1)$-simplices, where $(n+1)$ is the space-time dimension. These
can be glued in $(f-1) = (n+1)$ ways to a given boundary triangulation, where
$f$ is the number of faces of the simplex.  These moves will be
identified with the action of the local evolution operator and in this
way we can again interpret the projector as a sum over
triangulations. This is described in more detail for (2+1)d in the
next section.

However in higher dimensions
boundary states no longer only depend on the number of boundary
simplices but also their connectivity, which is related in a 
complicated way to the limiting continuum geometry of a discrete state. This
makes it very difficult to derive the action of a Hamiltonian
differential operator.
 These issues are
currently under investigation

\section{Spin-foam models and loop quantum gravity}

We will now discuss spin-foam models of $BF$ theory and quantum
gravity.  In general, a spin-foam is a 2-dimensional complex with
faces labelled by representations and edges labelled by intertwining
operators.  These structures were first used in path-integrals for
quantum gravity by Reisenberger~\cite{Reisenberger:1994aw} and now
provide a unifying framework for many approaches to quantum gravity
(see~\cite{Oriti:2001qu} and~\cite{Baez:1999sr} for reviews and
references).

In order to make contact with our discussion of dynamical
triangulations, we will make use of the fact that the dual of a
spin-foam is a triangulation of space-time with labelled edges. We
will see how one can provide a definition of the projector (not sum
over geometries) in terms of a sum over such space-time
triangulations. In contrast to the dynamical triangulations framework,
however, we also sum over the labellings of the edges in this approach.
We will show how a new regularisation of the projector will lead
naturally to this class of models.  The focus will be on the (2+1)d
case, and we indicate at the end how the results can be generalised.

In 2+1 dimensions general relativity is equivalent to a topological
$BF$ theory due to the absence of local degrees of freedom or
gravitons.  In the Euclidean case the fundamental variables are an
su(2) Lie algebra-valued space-time connection $\bA$ and an
su(2)-valued one form $\bB$.  The action is given by:
\begin{equation} \label{BF-action}
S[\bB,\bA] = \int_{\Man} \mathrm{Tr}[\bB\wedge F(\bA)],
\end{equation}
where $F(\bA)$ is
 the curvature of $\bA$.
By restricting to space-time manifolds of the form $\man = \Sigma \times
\Bbb{R}$, where $\Sigma$ is a compact 2-dimensional spatial manifold, we can
pass to a Hamiltonian formulation. The configuration space variable is
given by the restriction
$A$ of the connection $\bA$ to $\Sigma$. The Hamiltonian is a sum
of two constraints\footnote{Note that there is not need for a separate
2-diffeomorphism constraint as such transformations are already
generated by the curvature constraint.}.  The first imposes
$SU(2)$-gauge invariance and the second, also referred to as the
curvature constraint, implies that $F(A) = 0$, i.e. that the
connection is flat.  A good overview and a comprehensive list of
references can be found in~\cite{Carlip98}.

It can be shown that the above is equivalent to a Chern-Simons gauge
theory which was first used by Witten in~\cite{Witten:1988hc} to
obtain a quantisation of 2+1 gravity.  Here we will be using the
discrete Ponzano-Regge~\cite{Ponzano68} or Tuarev-Viro~\cite{Turaev:1992hq}
approaches to quantisation as they will resemble the dynamical
triangulations models discussed above.

\subsection{Quantum theory}

The kinematical framework for the Ponzano-Regge model can be described 
by loop
quantum gravity as first noticed by Rovelli~\cite{Rovelli:1993kc}.
The kinematical Hilbert space $\Hk$ is constructed from a space of
gauge-invariant wave-functions on the configuration space of
connections. This space is spanned by the so-called
spin-networks~\cite{Rovelli:1995ac}, which are the natural
generalisations of the Wilson loop. 

Spin-networks are functions of connections with support on graphs
embedded in space, which in 2+1 we choose to be trivalent.  More
precisely, given any closed, oriented, trivalent graph $\Gamma$ with
$n$ edges embedded in $\Sigma$ and a labelling of the edges by
representations of SU(2) denoted by $\vj = \{j_1,...,j_n\}$, we can
construct a spin-network function\footnote{Note that because we have
chosen graphs to be trivalent we do not require a choice of
intertwiner for the vertices. In 2+1 this does not imply a loss of
generality since the theory can be solved exactly in terms of these
spin-networks. Whether this is also possible in 3+1 dimensions with 4
valent graphs is unclear.} $\Psi_{\Gamma,\vj}(A)$ as follows.  Let us
denote the holonomy of the connection $A$ along the edge $e \in
\Gamma$ by $U_e[A]$. This can be thought of as an SU(2) group element.
 Thus for every edge $e$ labelled by a spin $j_e$ we can
 construct a matrix $\rho_{j_e}(U_e[A])^a_b$, where
 $\rho_{j_e}$ is the spin-$j_e$ representation of SU(2). We can associate
 the  index $a$ with the vertex to which the edge $e$ is directed and
 the index $b$ to the vertex at the other end of $e$. Taking the
 tensor product of these matrices gives us a tensor with three indices
 for every vertex in $\Gamma$. We can contract these indices with the unique
 invariant tensor in the tensor product of the representations on the
 edges incident on the vertex. This  gives us the gauge-invariant function
 $\Psi_{\Gamma,\vj}(A)$.

An inner product can now be imposed on the space generated by these
functions by demanding that spin-networks be orthonormal,
i.e.:
\begin{equation}
\ipk{\Psi_{\Gamma,\vj}}{\Psi'_{\Gamma',\vj'}} =
\delta_{\Gamma,\Gamma'} \delta_{\vj,\vj'}.
\end{equation} 
Completion with respect to this inner product gives us the kinematical 
Hilbert space $\Hk$. 

We can also describe the trivalent
spin-networks in terms of their dual triangulations $\bd$, making the
link to earlier discussions. Each
edge of $\bd$ is labelled
by the spin on the edge in $\Gamma$ intersecting it. 
We will switch between these two perspectives whenever convenient and
spin-networks will be labelled either by graphs, triangulations or not 
at all.

To proceed we now need to implement the Hamiltonian or curvature
constraint $\Ham$ and reduce to the final, physical, Hilbert space $\Hp$. 
As before this is done by constructing the generalised projector.
In contrast to the previous section one can give an exact expression
of this projector in terms of sums over triangulations.

As shown by Ooguri~\cite{Ooguri:1992ni}, the Ponzano-Regge model
defines the projector for 2+1 quantum gravity. This is most easily
introduced in terms of triangulations $\Del$ of $\Man= \Sigma \times
[0,1]$. We will distinguish between the interior edges $\be$ and the
edges $e$ of the boundaries $\bd_0$ and $\bd_1$ of $\Del$.  All edges
are labelled by representations $j_e$ of SU(2).  Hence we can
associate to every tetrahedron \mbox{$T
\subset \Del $} a normalised 6-J symbol denoted by $T\{\vj_T\}$, where
$\vj_T$ is the set of spins on the boundary edges of $T$, which can
include edges that lie
on the boundary of $\Del$. To every triangulation $\Del$ with given
boundary $\bd_0$, $\bd_1$ we can now define an amplitude
$Z_\Del(\bd_0,\bd_1)$ as a sum over all possible labellings of the
interior edges that are compatible with the labelling of the boundary:
\begin{equation}
Z_\Del(\bd_0,\bd_1) := \prod_{e}\sqrt{2j_{e} +1} \sum_{j_\be} \prod_\be
(2j_\be+1)
 \prod_T T\{\vj_T\}. 
\end{equation}
This sum can be infinite and needs to be regulated. This can be done
by introducing a cut-off as is done by Ooguri in~\cite{Ooguri:1992ni}
by using quantum groups. The latter approach gives the Tuarev-Viro
model~\cite{Turaev:1992hq} which is related to quantum gravity with a
cosmological constant.  This will not concern us in the following and
we assume that some choice of regularisation has been made to make the
above expression well-defined.
 
Given two
spin-networks $\Psi_0$ and $\Psi_1$, dual to triangulations $\bd_0$ and 
$\bd_1$, we can now define the physical inner
product as a sum over triangulations $\Del$ that have boundary $\bd_0+\bd_1$:
\begin{equation}
\ipk{\Psi_1}{\Po\Psi_0} = \sum_{\Del} Z_{\Del}(\bd_0,\bd_1)
\end{equation}
Using identities between the 6-J symbols one can show that
$Z_\Del(\bd_0,\bd_1)$ is in fact independent of the triangulation
$\Del$ and
depends only on $\bd_0$ and $\bd_1$.  Hence the matrix elements of
$\Po$ are given by $\infty \times Z(\bd_0,\bd_1)$, where $Z$ is
evaluated using any chosen triangulation with the correct boundary
data. Such an infinite factor is a common feature in the RAQ procedure
whenever the constraints do not form the Lie-algebra of a compact
group\footnote{Recall that  the projector averages over the action
generated by the Hamiltonian constraint.
}~\cite{Gomberoff:2000gr}. The factor can be dropped by redefining
the physical inner product and we have the final result\footnote{The
reason that we chose to include a sum over triangulations initially is
that it will play an important role in the following. The above
procedure is also necessary for gravity in higher dimensions
where we no longer have
triangulation independence.}:
\begin{equation}
\ipp{\tPsi_1}{\tPsi_0} = Z(\bd_0,\bd_1).
\end{equation}

A crucial property for the identification of the amplitude $Z$ with a
projector is the property:
\begin{equation}\label{projectionprop}
\sum_j Z(\bd_1,\bd_j)Z(\bd_j,\bd_0) = Z(\bd_1,\bd_0),
\end{equation}
where the sum is over all colourings $\bd_j$ of an intermediate
triangulation $\bd$.

As has been shown by Ooguri~\cite{Ooguri:1992ni}, this definition of
$\Po$ leads to a well-defined theory of quantum gravity, which is
equivalent to Witten's Chern-Simons approach.  If we formulate the
above in terms of spin-networks instead of their dual triangulations
we obtain the spin-foam model of 2+1 loop quantum gravity.  A
spin-foam is simply the 2-skeleton of the dual 2-complex of a given
triangulation.

The sum over spins on a triangulation can be given an interpretation
as a sum over geometries or path integral in several ways. Originally
the Ponzano-Regge model was an ad hoc modification of the Regge model
of gravity. Here one describes (piecewise linear) metrics by an
assignment of edge lengths and angles to the tetrahedra in a
triangulation of space-time. By restricting edge lengths to take on
only integer and half-integer values we obtain the Ponzano-Regge
model.

The actions of the two models can be related in a semi-classical limit
where we take spins $j$ to be large. In this limit the amplitude for a
single tetrahedron given by the 6-J symbol can expressed as:
\begin{equation} \label{semi-class}
T\{\vj_T\} \sim (6\pi V_T)^{-1/2} (e^{-i(S_R +\pi/4)} +
e^{i(S_R+\pi/4)}),
\end{equation}
where $V_T$ is the volume of the tetrahedron sides of length $L =
\frac{1}{2}(j +\frac{1}{2})$ and $S_R$ is the Regge action.
 Hence it is natural to identify spins with geodesic lengths in the
 Ponzano-Regge model. The fact that the 6-J symbol describes a sum of
 two exponentials with opposite signs is related to the fact that the
 Ponzano-Regge model defines a true projector and hence not a sum over
 geometries as we comment on again below.

A perhaps better understanding of the origin of the discreteness in
the edge lengths comes via the spin foam approach. It is possible to
show that Ponzano-Regge model arises if one discretises the
path integral:
\begin{equation}
\int [d\bB][d\bA] e^{-iS[\bB,\bA]},
\end{equation}
where the action is defined in equation~(\ref{BF-action}). The spins
labelling edges then arise after an integration over $\bB$, as the
Fourier components of the connection $\bA$ restricted to the edges of
the triangulation. Furthermore, one can construct a length operator for
loop quantum gravity and one finds that the spins are precisely proportional
to the quantised edge lengths as first noticed by Rovelli
in~\cite{Rovelli:1993kc}.

Since  spin-foam models involve a sum over edge lengths, a
continuum limit cannot be taken in the same way as in the dynamical
triangulations models where edge lengths are taken to 0. The
implications of this will be discussed below. It turns out that we
need to take the scale of the edge lengths to 0. Note however that because
the special cases of  $BF$ theory and 2+1 quantum gravity are topological these
models can be defined exactly in terms of a single triangulation
without taking a continuum limit as described above.

Note also that the sum over geometries that is described by the Ponzano-Regge
and more general spin-foam models is not gauge fixed. In other words
different spin assignments of a triangulation can correspond to the
same geometry. The rationale behind the spin foam
approach is that instead of a picking a representative of each gauge
equivalence class and integrating only  over these we are in
effect averaging over the equivalence classes. This will be important later.

\subsection{The projector and elementary moves}

 In order to derive the sum over triangulations picture from the
formal definition of the projector we
 now describe how one can construct $Z_\Del(\bd_0,\bd_1)$ for any
$\Del$ and $\bd_1$ given an initial triangulation $\bd_0$ by an
iterative sequence of elementary moves. These are the Pachner moves in
2+1 dimensions by which all triangulations can be related.

\begin{figure}
\begin{center}
\includegraphics*[width=3in,keepaspectratio]{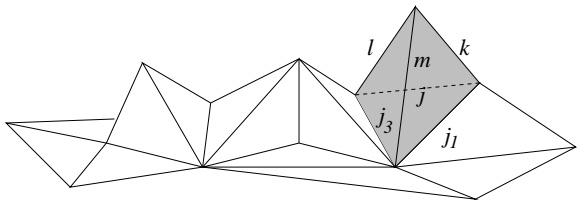}
\caption{Gluing one face a tetrahedron to $\bd_0$. Only the relevant
spins are labelled.}\label{glue1}
\end{center}
\end{figure}

\begin{figure}
\begin{center}
\includegraphics*[width=3in,keepaspectratio]{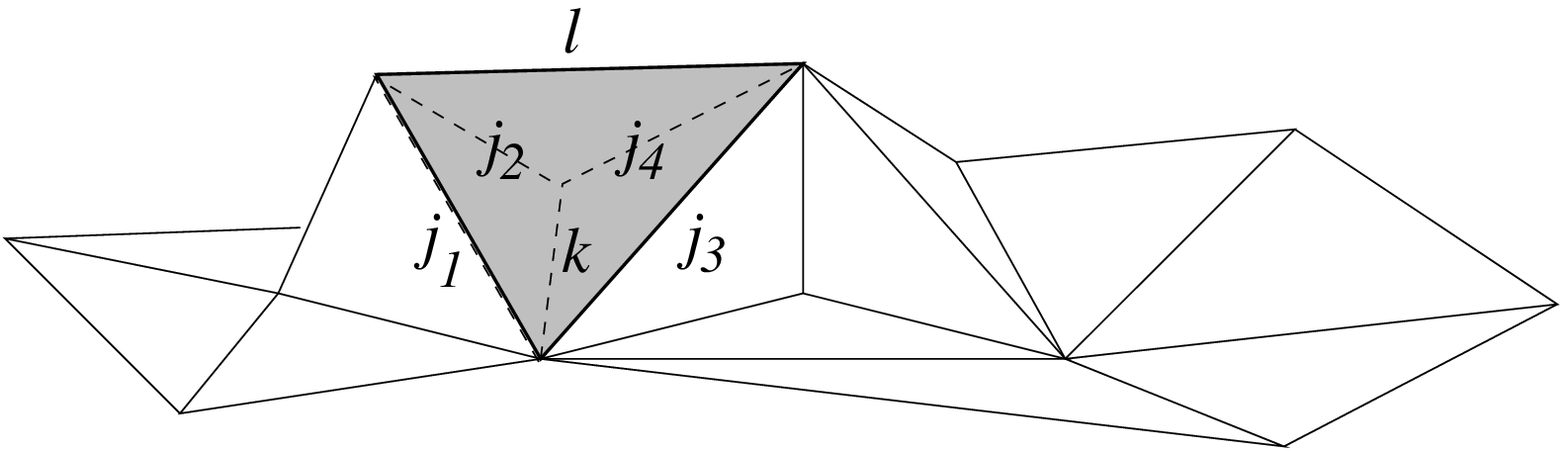}
\caption{Gluing two faces of a tetrahedron to $\bd_0$.}\label{glue2}
\end{center}
\end{figure}

\begin{figure}
\begin{center}
\includegraphics*[width=3in,keepaspectratio]{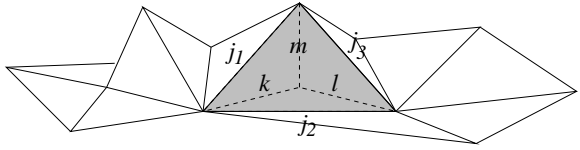}
\caption{Gluing 3 faces of a tetrahedron to $\bd_0$.}\label{glue3}
\end{center}
\end{figure}

There are 3 ways of gluing a tetrahedron $T$ onto a given triangulation
$\bd_0$ of $\Sigma$. These are shown in figures~\ref{glue1},
\ref{glue2} and \ref{glue3}: 
\begin{enumerate}
\item We can attach one face of $T$ to a triangle in $\bd_0$. To this
tetrahedron we associate the factor 
\begin{equation}\label{g1}
 T\left\{ \begin{array}{ccc} 
 j_1 & j_2 & j_3 \\ l & m & k    \end{array}\right\} 
\sqrt{(2k+1)(2l+1)(2m+1)},
\end{equation}
 where we have written out the 6-J symbol and $k,l,m$ are the spins on the edges
of $T$ that are not glued to $\bd_0$

\item We can attach two faces of $T$ to $\bd_0$. This is associated
with a factor 
\begin{equation}\label{g2}
T\left\{\begin{array}{ccc} 
j_1 & j_3 & l \\ j_4 & j_2 & k \end{array}\right\}
 \sqrt{(2k+1)(2l+1)}
\end{equation}
 where $k$ labels the
edge that lies in the interior of the two triangles glued to $\bd_0$
and $l$ is the spin on the opposite edge.

\item We can glue three faces of $T$ to $\bd_0$. In this case we have
a factor of 
\begin{equation}\label{g3}
T\left\{\begin{array}{ccc} 
j_1 & j_2 & j_3 \\ l & m & k \end{array}\right\}  
 \sqrt{(2k+1)(2l+1)(2m+1)}
\end{equation}
 where $k,l,m$ label
the edges in the interior of the three triangles attached to $\bd_0$.
\end{enumerate} 
 \begin{figure}
\begin{center}
\includegraphics*[width=3in,keepaspectratio]{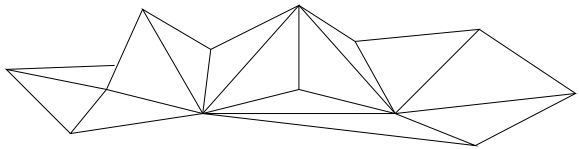}
\caption{A boundary triangulation $\bd_0$.}\label{base}
\end{center}
\end{figure}

Once we have completed the gluing\footnote{In terms of spin-foams we
are gluing together elementary atoms as is described
in~\cite{Reisenberger:2000zc}. The different ways of gluing triangles
correspond to the choice of which boundary edges of the atom we wish
to glue to the given spin-network. } we obtain a new boundary
triangulation $\bd'$. Using this boundary we can repeat the above
procedure. In this way we can construct any triangulation $\Del$ with
any boundary $\bd_1$ given an original boundary $\bd_0$.  By taking
the product of the factors associated to the tetrahedra we recover the
amplitude $Z_\Del(\bd_0,\bd_1)$.

Equivalently such a procedure can be viewed as constructing a sequence
of spin-networks via a sequence of operator actions, which can be
formalised as follows. We introduce the 3 move operators $\m_1$,
$\m_2$ and $\m_3$, that act locally on spin-networks:
\begin{figure}[h]\label{moves}
\begin{center}
\begin{tabular}{rcl}
$\m_1$
\parbox[c]{0.6in}{\includegraphics*[width=0.6in,keepaspectratio]{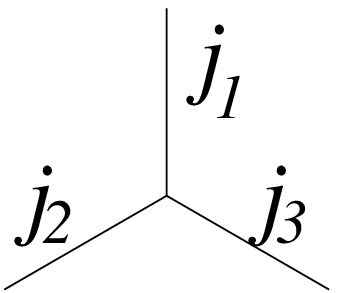}}
&=& 
$\sum_{k,l,m} T\left\{ \begin{array}{ccc} 
 j_1 & j_2 & j_3 \\ l & m & k    \end{array}\right\} 
\sqrt{(2k+1)(2l+1)(2m+1)}$
\parbox[c]{0.6in}{\includegraphics*[width=0.6in,keepaspectratio]{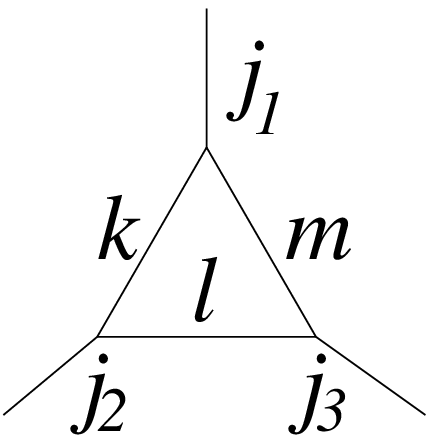}}
\\
\vspace{0.2cm}
\\
$\m_2$
\parbox[c]{0.6in}{\includegraphics*[height=0.6in,keepaspectratio]{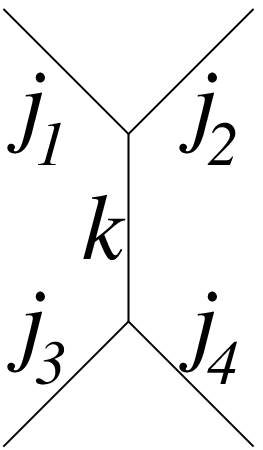}}
&=&
$\sum_l T\left\{\begin{array}{ccc} 
j_1 & j_3 & l \\ j_4 & j_2 & k \end{array}\right\}
 \sqrt{(2k+1)(2l+1)}$\ \ 
\parbox[c]{0.5in}{\includegraphics*[width=0.6in,keepaspectratio]{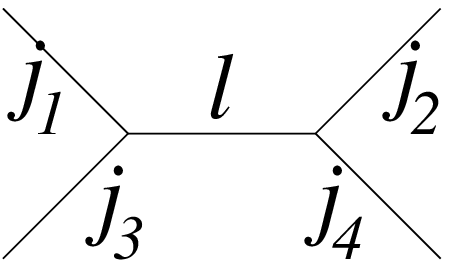}}\\
\vspace{0.2cm}
\\
$\m_3$
\parbox[c]{0.6in}{\includegraphics*[width=0.6in,keepaspectratio]{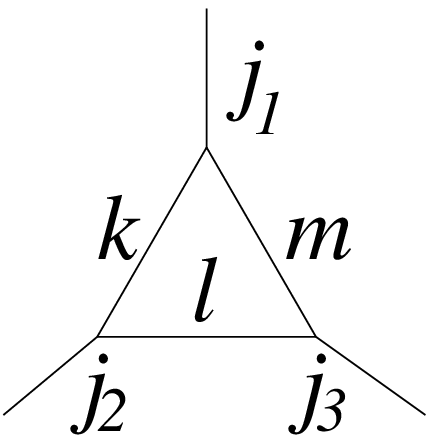}}
 &=&
$T\left\{\begin{array}{ccc} 
j_1 & j_2 & j_3 \\ l & m & k \end{array}\right\}  
 \sqrt{(2k+1)(2l+1)(2m+1)}$
\parbox[c]{0.6in}{\includegraphics*[width=0.6in,keepaspectratio]{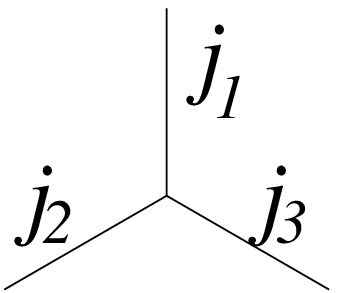}}

\end{tabular}
\end{center}
\end{figure}

\begin{figure}
\begin{center}
\includegraphics*[height=1in,keepaspectratio]{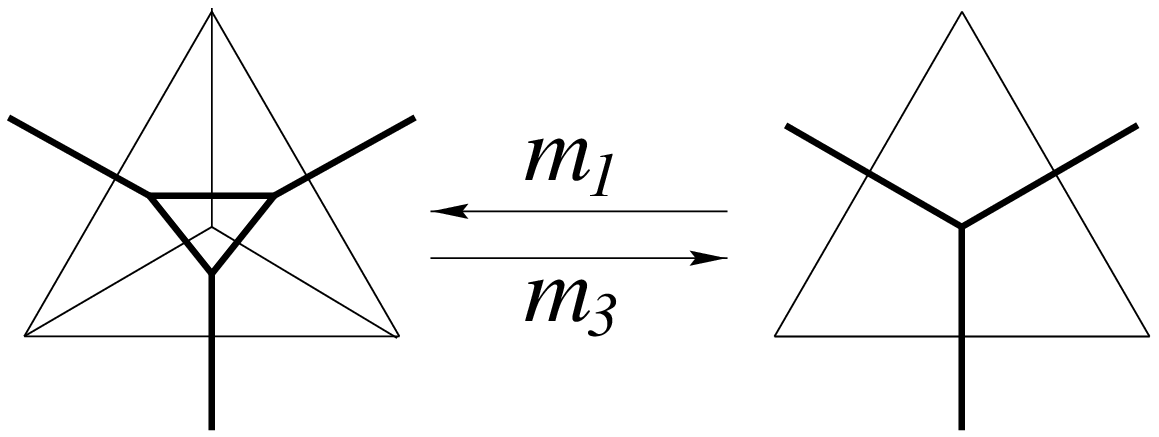} \\
\includegraphics*[height=1in,keepaspectratio]{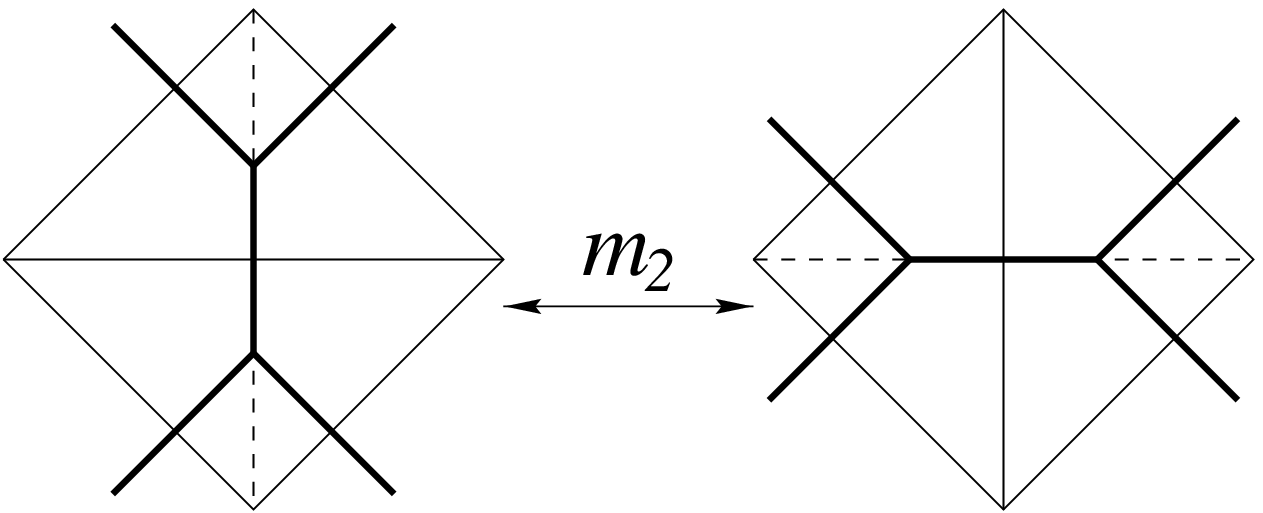}
\caption{The action of the elementary moves on spin networks are dual
to the addition of tetrahedra. Tetrahedra are depicted by the thin
lines and are viewed from the top. For example, we can view top left
figure as boundary triangulation consisting of 3 triangles with its
dual graph. By gluing a tetrahedron to all three triangles we obtain a
new boundary triangulation consisting of just one triangle as shown on
the right. This corresponds to the move $\m_3$ acting on the dual
graphs.}\label{duality}

\end{center}
\end{figure}

By a finite sequence of such moves one can relate any two trivalent
spin-networks. Note that $\m_1$, $\m_2$ and $\m_3$ are precisely dual
to the three possible gluings of tetrahedra as shown in
figure~\ref{duality}. The amplitudes also correspond to those given in
equations~(\ref{g1}), (\ref{g2}) and~(\ref{g3}).

 Hence any coloured triangulation $\Del$ can be
viewed as a given sequence of such elementary moves.

As with the dynamical triangulations models, we wish to use the fact
that the projector can be expressed as an ordered sequence of
operations in order to derive the above action from the formal expression
of the projector. This will clarify how the elementary move operators
$\m_1$, $\m_2$, $\m_3$ are related to evolution operators and in
consequence the Hamiltonian.

\subsection{Regularisation of the projector II}

We wish to provide a regularisation of the generalised projector as a
product of elementary local operators in analogy to the approach in
section~\ref{expansion1}. These elementary operators are to be
identified with 
the moves $\m_1$, $\m_2$ and $\m_3$. But note that in contrast to the
dynamical triangulations case the these moves do not involve a fixed
proper time, instead the describe a sum over all possible proper
times. This needs to be reflected in our expansion.

It turns out that the correct starting point for our calculation is
the following expression for the projector:
\begin{equation} \label{new-projector}
\Pro := \int [dN(x,t)] e^{-i\int_0^1 dt \int d^2x N(x,t) \Ham(x)},
\end{equation}
where we are integrating over all lapse functions on the cylinder
$\Sigma \times [0,1]$. The reason we need to extend the range of
integration as compared to our previous definition in equation~(\ref{projector})
is that the path integral that we wish to recover is not gauge fixed
as mentioned already above. 
This is reflected in the fact that the two definitions of the projector
are related by:
\begin{equation}
\int [dN(x,t)] e^{-i\int_0^1 dt \int d^2x N(x,t) \Ham(x)} = \N
\int[dT(x)] e^{-i\Ham[T]},
\end{equation}
where $T(x) = \int_0^1 dt N(x,t)$ and $\N$ is an infinite normalisation
factor, which should correspond to the volume of the gauge equivalence 
classes.
Finally, we note that in this section we integrate over the entire
range of lapse functions since we want to describe a genuine projector.

Let us now proceed with the regularisation of $\Pro$ by replacing
integrals by Riemann sums. First we consider the integral over the
coordinate time $t$ and  we replace the integration over functions
$N(x,t)$ by a product of integrations over spatial functions $N(x)$.
This leads to the following regularised expression:
\begin{eqnarray}
\Pro_\e &=& \int [dN(x,t)] e^{-i\sum_{t=0}^k \e \int d^2x N(x,t\e) \Ham(x)}\\
&=& \prod_{t=0}^k \int [dN(x)] e^{-i\e\int d^2x N(x) \Ham(x)},
\end{eqnarray}
where $k = 1/\e$.
Now we proceed analogously with the remaining spatial integral in the
exponential. To do this we need to introduce a square lattice with
a set of vertices $V$ as before.
\begin{eqnarray}
\Proreg &=& \prod_{t=0}^k \int [dN(x)] e^{-i\e\sum_{x\in V} l^2  N(x)
\Ham(x)}\\
&=& \prod_{t=0}^k \prod_{x\in V} \int_{-\infty}^{\infty} dT e^{-i\e
l^2 T \Ham(x)} \\
&=&  \prod_{t=0}^k \prod_{x\in V} \int_{0}^{\infty} dT \left( e^{-i\e
l^2 T \Ham(x)} +e^{i\e
l^2 T \Ham(x)} \right),
\end{eqnarray}
where we are left with a regular integral over $T \in \Bbb{R}$.
Let us now set
\begin{equation} \label{new-evolution}
\Ureg(x) := \int_{0}^{\infty} dT \left( e^{-i\e
l^2 T \Ham(x)} +e^{i\e l^2 T \Ham(x)} \right)
\end{equation}
then we have the final expression:
\begin{equation}\label{new-expansion}
\Proreg = \prod_{t=0}^k \prod_{x\in V} \Ureg(x)
\end{equation}

This expression is similar to equation~(\ref{expansion}) and can be
interpreted as a sum over triangulations. There are however essential
differences, most notably in the definition of $\Ureg$. Let us take
these in turn.

\subsection{Interpretation}

The operator $\Ureg(x)$ we have defined above is given by a sum over
local evolution operators for all proper times. Hence it is natural to 
make the identification:
\begin{equation}
\Ureg(x) = \m_1 +\m_2 +\m_3
\end{equation}
The argument $x$ of the operator $\Ureg(x)$ determines the
location of the action of the move operators $\m_i$ on  a given spin network.
The integral over $T$ in the definition of $\Ureg$ essentially reflects
the sum over spins in the definition of the elementary moves.
This identification is further supported by the fact that $\Ureg$
contains two terms involving an evolution in opposite proper time
directions. This mirrors the semi-classical interpretation of the 6-J
symbol given a sum of two exponentials of the Regge action described in 
equation~(\ref{semi-class}). Here we can see that this is a
result of the integration range in the integral over lapse functions in 
the definition of the projector and hence a consequence of the fact
that we are not dealing with a sum over geometries but a genuine
projection operator.

\begin{figure}
\begin{center}
\includegraphics*[height=1in,keepaspectratio]{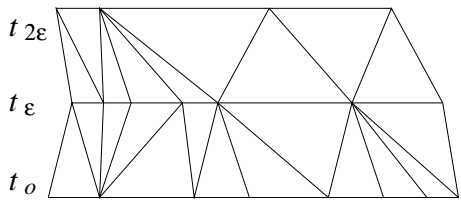}
\caption{A (1+1)d triangulation with a slicing structure.}\label{slices}
\end{center}
\end{figure}

Once we have made the identification of $\Ureg$ with the elementary
moves we can see how equation~(\ref{new-expansion}) describes a sum
over triangulations.  Each operator $\Ureg(x)$ adds a tetrahedron in
all possible ways and with all possible spins to a boundary
triangulation at location $x$. As we take the limit $l \ra 0$ the
product over positions $x\in V$ then describes the addition of
tetrahedra to every position on the initial boundary, such that we
obtain a single layer of tetrahedra.  We then repeat
this at the next co-ordinate time step.  The result is that we obtain
a sum over triangulations which each have a layered structure as
depicted in figure~\ref{slices}.

These layers can be interpreted as the slices $\{t\} \times \Sigma$ in
the foliation  $\Sigma \times [0,1]$. The spins on the edges of the
tetrahedra describe the multi-fingered proper-time separation between
the slices. The sum over spins then includes all possible proper-time
separations between initial and final boundaries.

We have so far not addressed how the parameter $\e$ is related to the
moves $\m_i$ and how this continuum limit $\e \ra 0$ is to be
interpreted. Indeed, as mentioned above we can no longer identify $\e$ 
simply with a proper time separation since we are summing over all
possible proper times for every tetrahedron. Instead we see from
equation~(\ref{new-evolution}) that $\e$ multiplies the proper time $T$ 
and hence is responsible for determining the scale of the sum over
proper times.
This indicates that $\e$ should be identified with the scale
associated in the relation between spins and edge lengths of
tetrahedra. Indeed, in a continuum limit we would like the edge
lengths to become small in some units. 
Hence, we should set physical lengths proportional to spins via: $L \sim 
\e j$.

Remarkably this parameter arises naturally in 3+1 loop quantum
gravity. It is the Immirzi
parameter~\cite{Immirzi:1996dr,Rovelli:1998na} $\gamma$ , which
describes a one-parameter family of classical canonical
transformations which in turn give rise to a one-parameter family of
\emph{inequivalent} quantum theories.  This parameter relates spin labels of
spin networks to the spectra of physical geometrical operators. It
determines that the eigenvalues are given in units of $(\gamma l_P)^a$
where $l_p$ is the Planck length and the power $a$ is the
dimension of the geometric observable we are considering.  Hence, we
deduce that we should make the identification $\gamma =
\e$ and a continuum limit of loop quantum gravity is obtained by
taking the Immirzi parameter to 0. This is especially interesting as
it has been proposed in an entirely different context by
Bojowald~\cite{Bojowald:2001ep}, 
who studied reduced cosmological models using loop quantum gravity.

We note again that $BF$ theory and (2+1)d gravity are special in that they 
are topological and the limit $\e \ra 0$ does not in fact need to be taken to 
define the theory. This can be interpreted as stating that it does not 
matter at which scale we approximate a continuum geometry. Since there 
are only a finite number of degrees of freedom any discrete
approximation will capture the full physical content of the theory.

\subsection{The Hamiltonian}

In the previous section we have provided a well motivated expansion of 
the projector describing the Ponzano-Regge sum over triangulations.
Some of the attractive features of this approach were:
\begin{itemize}
\item The natural identification of the sum over spins as a sum over
proper-times;

\item The explanation of the semi-classical limit of 6-J symbols in
equation~(\ref{semi-class}) as a result of the integration range in
the generalised projector;

\item The identification of the continuum limit as a limit in which
the scales of edge lengths are taken to 0;

\item The reflection of the fact that the Ponzano-Regge model does not 
describe a gauge fixed sum over metrics in the definition of the
projector given in equation~(\ref{new-projector}).
\end{itemize}

In this section we discuss a further motivation for our approach,
which is closely linked to questions concerning the Hamiltonian.
The point is that the local move
operators $\m_i$ leave physical states invariant, i.e.:
\begin{equation}\label{invar}
\ipp{\tPsi}{\m_1\tPhi} = \ipp{\tPsi}{\m_2\tPhi}=\ipp{\tPsi}{\m_3\tPhi} =
\ipp{\tPsi}{\tPhi}
\end{equation}
for all $\tPsi$, $\tPhi$
This is shown by Ooguri in~\cite{Ooguri:1992ni}.
Let us take $\m_2$ as an example. We have:
\begin{equation}
\m_2\Phi_{\bd_0} = \sum_{k} T\{\vj_T\}\sqrt{(2k+1)(2j+1)} \Phi_{\bd_k} 
= \sum_k Z(\bd_k,\bd_0) \Phi_{\bd_k},
\end{equation}
 Here $\Phi_{\bd_0}$ is the
spin-network dual to the coloured triangulation $\bd_0$ and $\bd_k$ is 
the triangulation which is obtained from $\bd_0$ by adding a tetrahedron
with external spin $k$ as shown in figure~\ref{glue2}. 
The last equality in the above equation follows from the definition of 
$Z$ and the fact that, due to the triangulation invariance of the model, 
the transition amplitude between the states
$\Phi_{\bd_0}$ and $\Phi_{\bd_l}$ is given by the amplitude of a
single tetrahedron, which in turn is the same as the amplitude of the
$\m_2$ move.

Hence,  it follows:
\begin{eqnarray}
\ipp{\tPsi_{\bd_1}}{\m_2\tPhi_{\bd_0}} &=& \sum_k Z(\bd_1,\bd_k)
Z(\bd_k, \bd_0) \\ 
&=& Z(\bd_1,\bd_0) = \ipp{\tPsi}{\tPhi},
\end{eqnarray}
where we have used the projection property~(\ref{projectionprop}) of
the amplitude $Z$.

This can also be understood as follows.  The constraints of 2+1
general relativity impose the condition that the connection is flat.
Physical states can be though of spin-network functions evaluated on
flat connections. But it can be shown that the action of the
elementary moves on spin-network functions restricted to flat
connections is the identity.

Physical states are orbits of kinematical states under the
exponentiated action of the Hamiltonian constraint. Hence
 the elementary moves map between states
in $\Hk$ that 
are related by such action and they can thus be viewed as elementary
time-evolutions. Thus it is natural to identify the moves with the
evolution operator $\Ureg$.

Note, however, that this is in contrast to earlier assumptions on the
action of the Hamiltonian in loop quantum gravity.  Let us recall that
in the Reisenberger-Rovelli approach the exponential in the projector
is expanded as follows:
\begin{equation}
\int [dT(x)] e^{-i\Ham[T]} = \int [dT(x)] \sum_{t=0}^{\infty}
\frac{(-i)^t}{t!} \Ham[T]^t,
\end{equation}
The idea was to identify $\Ham[T]$ with the total move operator 
\begin{equation}\label{tot-moves}
\Ham[T] = \EV :=
\int_\Sigma dx (a_1\m_1 + a_2\m_2 + a_3\m_3),
\end{equation}
where $a_1,a_2$ and $a_3$ are arbitrary coefficients that need to be
fixed.  This was motivated because the action of Thiemann's
Hamiltonian~\cite{Thiemann:1998aw} in 3+1 dimensions resembled the
move $\m_1$ albeit with a different amplitude.  The other moves need
to be included in the Hamiltonian if the multiple actions of the
Hamiltonian are to give rise to all triangulations. This is because
the integral over the scalar lapse functions does not alter the
topological action of the operators.  The search for such a
Hamiltonian, also referred to as crossing symmetric, has met only with
partial success~\cite{Gaul:2000ba}. In particular it is not clear how
the move $\m_2$, which is thought to be crucial for long range
propagation of excitations~\cite{Smolin:1996fz}, can be derived from
the quantisation of a classical expression of a Hamiltonian.

But from the discussion above one can understand why this might be the
wrong approach. Since the local moves leave physical states invariant
it is impossible that any combination of them will
annihilate physical states as would be required if they were to
correspond to a Hamiltonian constraint. This is because the integrated
move operator will be a sum of elementary operators acting at all
possible locations in the triangulation underlying a spin-network.
Each single action leaves physical states invariant. This implies the
following for
the action of the Hamiltonian defined in equation~(\ref{tot-moves}) on
spin networks $\Psi$ and $\Phi$
\begin{eqnarray}
\tilde{H}[T]\tPsi[\Phi] &:=& \int_\Sigma dx \bra{\Phi}\Pro(a_1\m_1 + a_2\m_2 +a_3
\m_3)\ket{\Psi}\\
&=& (a_1 n_1(\Psi) + a_2 n_2(\Psi) + a_3 n_3(\Psi))\delta_{\Psi,\Phi},
\end{eqnarray}
where we have used equation~(\ref{invar}) and the orthogonality of
spin-network states in the kinematical inner product.
The coefficients $n_1$, $n_2$, $n_3$ count the number of times that
the operators $\m_1$, $\m_2$ and $\m_3$ respectively can act on the
state $\Psi$.
But since these coefficients are state
dependent and the coefficients $a,b,c$ are not, the action of
$\tilde{H}[T]$ cannot be
made to vanish on all physical states.

So from our perspective it seems clear that the difficulties in
generalising the definition of the loop quantum gravity Hamiltonian
stems from identifying the move operators with Hamiltonian actions and 
not
evolutions. This is in line with views
expressed by Markopoulou~\cite{Markopoulou:1997ri}, where evolution of spin
networks is described in terms of the Pachner moves listed above. In
so far as these models should reproduce quantum gravity in 2+1
dimensions our work gives them a precise motivation in terms of
canonical loop quantum gravity and shows that the amplitudes for these
transitions should be given by the 6-J symbols as described above.

The problem we face is to find a Hamiltonian which can reproduce the
elementary moves when exponentiated. This seems a difficult task. One
possibility is that it would be possible to quantise the action
generated by the classical Hamiltonian directly, in the same way that
it is only possible to solve the diffeomorphism constraint by
quantising the action generated by it.

Another option that is currently being explored is to make use of the
close connection to the classical formulation of 2+1 gravity in terms
of 't Hoofts lattice model~\cite{'tHooft:1993gk}. Here states are
described by trivalent graphs with labellings corresponding to edge
lengths. It can be shown that Hamiltonian evolution increases and
decreases the edge lengths. Crucially, edge lengths can evolve to zero
at which point a change in the topology of the graph can occur. These
possible changes have been tabulated in~\cite{'tHooft:1993gz} and, in
the absence of matter, they turn out to be precisely the three
elementary evolution moves. Hence it is possible that Hamiltonian
action should act on edges by increasing and decreasing the values of
the spins by adding loops of spin 1 to all closed loops in the graph
of a given spin-network.

\subsection{Generalisations and outlook}

Our results hold for $BF$ theories in any dimension, which are all
topological field theories. 
These models are also expressed in terms of a single triangulation, where
the 6-J symbols are
replaced by their higher dimensional analogues. 

The situation looks more complicated for quantum gravity since in
space-time dimension higher than 3 the theory has an infinite number
of degrees of freedom, and is no longer equivalent to $BF$ theory and
no longer topological. However there still exist spin-foam models that
attempt to define a path integral for the theory both for Lorentzian
and Euclidean signatures (see~\cite{Oriti:2001qu} for a review).  In
this case the transition amplitudes are no longer triangulation
independent and we need to include the sum over triangulations to
restore covariance. However, our results have shown that the sum over
triangulations should be restricted to triangulations with a slicing
structure. Such restricted sums have been studied in the context of
Lorentzian dynamical triangulation models~\cite{Ambjorn:2001cv} and there is
mounting evidence that they have better behaviour than sums over
generic triangulations.

In addition, as soon as spin-foam models are no longer topological,
taking a continuum limit will become important. Our work indicates
that in loop quantum gravity and related spin-foam models this limit
involves considering large spins while taking the Immirzi parameter
$\gamma$ to 0.  This has been suggested independently by Bojowald in
the context of cosmological models~\cite{Bojowald:2001ep}. In
addition, the study of the geometric operators area, volume and angle
around a vertex of a spin-network carried out by Major and
Seifert~\cite{Major:2001zg} has revealed that in order to approximate
continuum geometries resembling what we observe, we need to consider a
large spin limit.

We note also that for non-topological theories we can no longer argue
that the elementary moves leave physical states invariant.
Presumably again this is linked to the need to take the continuum
limit in defining the evolution operators.
The physical consequences of this deserves to be investigated in more
detail.

\section*{Acknowledgements}

We are grateful to Jan Ambj{\o}rn for continuing support throughout this
work,  especially for the idea on how to derive the Hamiltonian
constraint in (1+1) dynamical triangulations. We would also like to
thank Marcus Gaul, B.\ Durhuus and Carlo Rovelli for helpful discussions.
This work was supported by the 
EU-network on ``Discrete Random Geometries'' grant HPRN-CT-1999-00161.

\section*{Appendix}

 We show that if a triangulation is generated by the action of an
ordered product $\Urega(x_1)\cdots\Urega(x_s)$ acting on an auxiliary
state, then this is the unique
sequence of operators that will produce this triangulation, where
$\Urega(x)$ is defined in equation~(\ref{U-def}).
 We proceed
by induction over the number of triangles in a triangulation. 

Clearly the addition of one triangle to a boundary triangulation
corresponds to a unique action of on $\Urega(x)$. The position $x$ is
given by the co-ordinate of the node on the right edge of the
triangle glued to the boundary.

Let us assume that any triangulation with $n$ triangles has a unique
sequence of evolution operators corresponding to it.
Now let us consider a triangulation with $n+1$ triangles, which can be
described by at least one sequence of operators. 
This triangulation will have a set of triangles that is attached to
the boundary. This set inherits an ordering from the order of the
coordinates of the nodes on the edges that the triangles are glued
to. Hence,  the triangulation 
will contain a unique triangle, which is defined by the
fact that it is attached to a
boundary edge 
with the largest value of the node coordinate. Since
the products of evolution operators in the expansion of $\Prorega$ are
ordered this triangle has to be constructed first. Hence the first
operator in any sequence of operators corresponding to our
triangulation  is fixed and unique. 

But now we need to add only $n$ more triangles to obtain our
triangulation. By assumption there is only one sequence of operators
that will describe this triangulation with $n$ triangles.

Hence we conclude that every triangulation with $n+1$ triangles has a
unique corresponding product of operators $\Urega(x_1)\cdots\Urega(x_{n+1})$. qed.


\begin{thebibliography}{999999}

\bibitem{Ambjorn:1999nc}
J.~Ambjorn, J.~Jurkiewicz, and R.~Loll.
\newblock Lorentzian and euclidean quantum gravity: Analytical and numerical
  results.
\newblock 1999, hep-th/0001124.

\bibitem{Ambjorn:2001cv}
J.~Ambjorn, J.~Jurkiewicz, and R.~Loll.
\newblock Dynamically triangulating lorentzian quantum gravity.
\newblock {\em Nucl. Phys.}, B610:347--382, 2001, hep-th/0105267.

\bibitem{Baez:1999sr}
John~C. Baez.
\newblock An introduction to spin foam models of bf theory and quantum gravity.
\newblock {\em Lect. Notes Phys.}, 543:25--94, 2000, gr-qc/9905087.

\bibitem{Bojowald:2001ep}
Martin Bojowald.
\newblock The semiclassical limit of loop quantum cosmology.
\newblock {\em Class. Quant. Grav.}, 18:L109--L116, 2001, gr-qc/0105113.

\bibitem{Carlip98}
Steven Carlip.
\newblock {\em Quantum gravity in 2+1 dimensions}.
\newblock Cambridge University Press, 1998.

\bibitem{Durhuus:2001sp}
B.~Durhuus and C.~W.~H. Lee.
\newblock A string bit hamiltonian approach to two-dimensional quantum gravity.
\newblock 2001, arXiv:hep-th/0108149.

\bibitem{Gaul:2000ba}
Marcus Gaul and Carlo Rovelli.
\newblock A generalized hamiltonian constraint operator in loop quantum gravity
  and its simplest euclidean matrix elements.
\newblock 2000, gr-qc/0011106.

\bibitem{Giulini:1999kc}
D.~Giulini.
\newblock Group averaging and refined algebraic quantization.
\newblock {\em Nucl. Phys. Proc. Suppl.}, 88:385, 2000, gr-qc/0003040.

\bibitem{Gomberoff:2000gr}
Andres Gomberoff.
\newblock On group averaging for non-compact groups.
\newblock 2000, hep-th/0012040.

\bibitem{Halliwell:1992nj}
Jonathan~J. Halliwell and Miguel~E. Ortiz.
\newblock Sum over histories origin of the composition laws of relativistic
  quantum mechanics.
\newblock {\em Phys. Rev.}, D48:748, 1993, gr-qc/9211004.

\bibitem{Immirzi:1996dr}
Giorgio Immirzi.
\newblock Quantum gravity and regge calculus.
\newblock {\em Nucl. Phys. Proc. Suppl.}, 57:65, 1997, gr-qc/9701052.

\bibitem{Klauder:2000gu}
John~R. Klauder.
\newblock Quantization of constrained systems.
\newblock {\em Lect. Notes Phys.}, 572:143--182, 2001, hep-th/0003297.

\bibitem{Major:2001zg}
Seth~A. Major and Michael~D. Seifert.
\newblock Modelling space with an atom of quantum geometry.
\newblock 2001, gr-qc/0109056.

\bibitem{Markopoulou:1997ri}
Fotini Markopoulou.
\newblock Dual formulation of spin network evolution.
\newblock 1997, gr-qc/9704013.

\bibitem{Marolf:2000iq}
Donald Marolf.
\newblock Group averaging and refined algebraic quantization: Where are we now?
\newblock 2000, gr-qc/0011112.

\bibitem{Ooguri:1992ni}
Hirosi Ooguri.
\newblock Partition functions and topology changing amplitudes in the 3-d
  lattice gravity of ponzano and regge.
\newblock {\em Nucl. Phys.}, B382:276--304, 1992, hep-th/9112072.

\bibitem{Oriti:2001qu}
Daniele Oriti.
\newblock Spacetime geometry from algebra: Spin foam models for non-
  perturbative quantum gravity.
\newblock 2001, gr-qc/0106091.

\bibitem{Ponzano68}
G.~Ponzano and T.~Regge.
\newblock Semiclassical limit of racah coefficients.
\newblock In F.~Bloch, editor, {\em Spectroscopic and Group Theoretical Methods
  in Physics}, New York, 1968. North-Holland.

\bibitem{Reisenberger:1994aw}
Michael~P. Reisenberger.
\newblock World sheet formulations of gauge theories and gravity.
\newblock 1994, arXiv:gr-qc/9412035.

\bibitem{Reisenberger:1997pu}
Michael~P Reisenberger and Carlo Rovelli.
\newblock *sum over surfaces* form of loop quantum gravity.
\newblock {\em Phys. Rev.}, D56:3490--3508, 1997, gr-qc/9612035.

\bibitem{Reisenberger:2000zc}
Michael~P. Reisenberger and Carlo Rovelli.
\newblock Spacetime as a feynman diagram: The connection formulation.
\newblock {\em Class. Quant. Grav.}, 18:121--140, 2001, gr-qc/0002095.

\bibitem{Rovelli:1993kc}
Carlo Rovelli.
\newblock The basis of the ponzano-regge-turaev-viro-ooguri quantum gravity
  model in the loop representation basis.
\newblock {\em Phys. Rev.}, D48:2702--2707, 1993, hep-th/9304164.

\bibitem{Rovelli:1998dx}
Carlo Rovelli.
\newblock The projector on physical states in loop quantum gravity.
\newblock {\em Phys. Rev.}, D59:104015, 1999, gr-qc/9806121.

\bibitem{Rovelli:1995ac}
Carlo Rovelli and Lee Smolin.
\newblock Spin networks and quantum gravity.
\newblock {\em Phys. Rev.}, D52:5743--5759, 1995, gr-qc/9505006.

\bibitem{Rovelli:1998na}
Carlo Rovelli and Thomas Thiemann.
\newblock The immirzi parameter in quantum general relativity.
\newblock {\em Phys. Rev.}, D57:1009--1014, 1998, gr-qc/9705059.

\bibitem{Smolin:1996fz}
Lee Smolin.
\newblock The classical limit and the form of the hamiltonian constraint in
  non-perturbative quantum general relativity.
\newblock 1996, gr-qc/9609034.

\bibitem{'tHooft:1993gk}
G.~'t~Hooft.
\newblock Classical n particle cosmology in (2+1)-dimensions.
\newblock In *'t Hooft, G. (ed.): Under the spell of the gauge principle*
  606-618, and Class. Quantum Grav. 10 (1993) Suppl., pp 79-91.

\bibitem{'tHooft:1993gz}
G.~'t~Hooft.
\newblock The evolution of gravitating point particles in (2+1)- dimensions.
\newblock {\em Class. Quant. Grav.}, 10:1023--1038, 1993.

\bibitem{Teitelboim:1982ua}
Claudio Teitelboim.
\newblock Quantum mechanics of the gravitational field.
\newblock {\em Phys. Rev.}, D25:3159, 1982.

\bibitem{Teitelboim:1983fh}
Claudio Teitelboim.
\newblock Causality versus gauge invariance in quantum gravity and
  supergravity.
\newblock {\em Phys. Rev. Lett.}, 50:705, 1983.

\bibitem{Teitelboim:1983fk}
Claudio Teitelboim.
\newblock The proper time gauge in quantum theory of gravitation.
\newblock {\em Phys. Rev.}, D28:297, 1983.

\bibitem{Thiemann:1998aw}
T.~Thiemann.
\newblock Quantum spin dynamics (qsd).
\newblock {\em Class. Quant. Grav.}, 15:839--873, 1998, gr-qc/9606089.

\bibitem{Turaev:1992hq}
V.~G. Turaev and O.~Y. Viro.
\newblock State sum invariants of 3 manifolds and quantum 6j symbols.
\newblock {\em Topology}, 31:865--902, 1992.

\bibitem{Witten:1988hc}
Edward Witten.
\newblock (2+1)-dimensional gravity as an exactly soluble system.
\newblock {\em Nucl. Phys.}, B311:46, 1988.

\end{thebibliography}

\end{document}